\documentclass[twocolumn]{aastex631}
\usepackage{multirow}
\usepackage{amsmath}
\usepackage{xcolor}
\usepackage{csquotes}
\usepackage{mathtools}
\usepackage{enumitem}
\usepackage{fancybox}
\usepackage{marvosym}
\usepackage{graphicx}
\usepackage{amssymb}
\usepackage{natbib, bm}
\usepackage{orcidlink}

\usepackage{tabularx}
\usepackage{booktabs}

\providecommand{\pgfsyspdfmark}[3]{}
\usepackage[english]{babel}

\newcommand{\mtwo}{M_{\rm{200}}}
\newcommand{\rtwo}{R_{\rm{200}}}

\newcommand{\mhalo}{{M}_{\rm{halo}}}

\newcommand{\mstar}{{M}_{\star}}

\newcommand{\msun}{{\rm M}_{\odot}}

\newcommand{\kms}{{\rm km \, s}^{-1}}

\newcommand{\lt}{<}
\newcommand{\gt}{>}
\newcommand{\z}{{\it z}}

\newcommand{\oii}{\hbox{\sc [O\,ii]}}

\DeclareRobustCommand{\ion}[2]{%
\relax\ifmmode
\ifx\testbx\f@series
{\mathbf{#1\,\mathsc{#2}}}\else
{\mathrm{#1\,\mathsc{#2}}}\fi
\else\textup{#1\,{\mdseries\textsc{#2}}}%
\fi}

\begin{document}

\title[The importance of gas starvation in driving satellite quenching in galaxy groups at \boldmath$z \sim 0.8$]
{The importance of gas starvation in driving satellite quenching in galaxy groups at \boldmath$z \sim 0.8$}

\correspondingauthor{Devontae Baxter}
\email{dcbaxter@ucsd.edu}

\author[0000-0002-8209-2783]{Devontae C. Baxter}
\altaffiliation{NSF Astronomy and Astrophysics Postdoctoral Fellow \\}
\affiliation{Department of Astronomy \& Astrophysics,
University of California, San Diego, 9500 Gilman Dr, La Jolla, CA 92093, USA \\}

\author[0000-0002-8425-0351]{Sean P. Fillingham}
\affiliation{Center for Cosmology, Department of Physics \& Astronomy,
University of California, Irvine, 4129 Reines Hall, Irvine, CA 92697, USA \\}

\author[0000-0002-2583-5894]{Alison L. Coil}
\affiliation{Department of Astronomy \& Astrophysics,
University of California, San Diego, 9500 Gilman Dr, La Jolla, CA 92093, USA \\}

\author[0000-0003-1371-6019]{Michael C. Cooper}
\affiliation{Center for Cosmology, Department of Physics \& Astronomy,
University of California, Irvine, 4129 Reines Hall, Irvine, CA 92697, USA \\}

\begin{abstract}

We present results from a Keck/DEIMOS survey to study satellite quenching in group environments at $z \sim 0.8$ within the Extended Groth Strip (EGS). We target $11$ groups in the EGS with extended X-ray emission. We obtain high-quality spectroscopic redshifts for group member candidates, extending to depths over an order of magnitude fainter than existing DEEP2/DEEP3 spectroscopy. This depth enables the first spectroscopic measurement of the satellite quiescent fraction down to stellar masses of $\sim 10^{9.5}~\msun$ at this redshift. By combining an infall-based environmental quenching model, constrained by the observed quiescent fractions, with infall histories of simulated groups from the IllustrisTNG100-1-Dark simulation, we estimate environmental quenching timescales ($\tau_{\mathrm{quench}}$) for the observed group population. At high stellar masses ($\mstar=10^{10.5}~\msun$) we find that $\tau_{\mathrm{quench}} = 2.4\substack{+0.2 \\ -0.2}$ Gyr, which is consistent with previous estimates at this epoch. At lower stellar masses ($\mstar=10^{9.5}~\msun$), we find that $\tau_{\mathrm{quench}}=3.1\substack{+0.5 \\ -0.4}$ Gyr, which is shorter than prior estimates from photometry-based investigations. These timescales are consistent with satellite quenching via starvation, provided the hot gas envelope of infalling satellites is not stripped away. We find that the evolution in the quenching timescale between $0 \lt z \lt 1$ aligns with the evolution in the dynamical time of the host halo and the total cold gas depletion time. This suggests that the doubling of the quenching timescale in groups since $z\sim1$ could be related to the dynamical evolution of groups or a decrease in quenching efficiency via starvation with decreasing redshift.

\end{abstract}

\keywords{Galaxy quenching (2040), Galaxy evolution (594), Galaxy groups (597), Galaxy environments (2029), Extragalactic astronomy (506)}

\section{Introduction} 
\label{sec:intro}

In exploring the influence of environment on the suppression (or \textquote{quenching}) of star formation, considerable attention has been given to galaxy clusters, the most massive and rarest galaxy overdensities \citep[e.g., see][]{Wetzel13, Muzzin14, Balogh16, Foltz18, Baxter23, Ahad24}. However, galaxy groups, which represent intermediate-density environments between the low-density field and galaxy clusters, are considerably more prevalent, with the majority of galaxies residing in groups \citep[e.g.,][]{Tully87, Eke04}. Given this, it is crucial to study galaxy groups across a broad range of redshifts and stellar masses, focusing on both the star-forming and quenched galaxy populations. Among other aspects, this will help to address major outstanding questions regarding the physical mechanisms responsible for the dramatic growth of the population of quenched galaxies since $z\sim2$ \citep[see][]{Bell04, Bundy06, Faber07}.


However, there exist numerous obstacles in studying group populations across cosmic time. Groups are far less galaxy-rich relative to clusters, making them less efficient to observe and more difficult to characterize. Furthermore, while there have been large spectroscopic surveys at $z\sim1$ (e.g., DEEP2 \citep{Newman13}, zCOSMOS \citep{Lily07}) that have identified hundreds of groups \citep{Gerke12}, they collectively lack the necessary depth, sampling density, and/or spectral resolution to accurately identify a substantial number of satellite galaxies. Additionally, these spectroscopic surveys are less complete for the crucial population of quenched galaxies due to selection biases favoring massive, star-forming galaxies.


A fruitful pathway forward to improving upon current group samples to $z\sim1$-- and the goal of this study -- is to perform targeted spectroscopic follow-up of faint group member candidates around spectroscopically-confirmed groups. These investigations are crucial to address the limitations of existing galaxy formation models, both semi-analytic and hydrodynamic, which struggle to reproduce observed satellite quenched fractions in groups and cluster environments \citep[e.g.,][]{Hirschmann14, DeLucia19, Xie20, Donnari21, Kukstas23}. 


One effective approach to improving existing models is to combine observations of group and cluster populations with $N$-body simulations to constrain the timescale over which satellite galaxies quench since first infall onto their host system. This timescale, commonly denoted $\tau_{\mathrm{quench}}$, is important as it encapsulates the efficiency at which star formation is quenched in dense environments and serves as a distinguishing feature between the various quenching mechanisms that are theorized to exclusively operate in galaxy groups and clusters \citep[e.g.][]{Wetzel14, Fham15, Wright19}. These include inefficient mechanisms like starvation \citep{Larson80} -- i.e., the slow consumption of cold gas in the absence of cosmological accretion -- that quench galaxies on timescales comparable to the cold gas depletion time, $\tau_{\mathrm{dep}} \equiv M_{\mathrm{gas}}/SFR$, where $M_{\mathrm{gas}}$ is the remaining cold gas mass and SFR is the star formation rate of the galaxy. In contrast, highly efficient mechanisms such as ram-pressure stripping \citep[RPS,][]{GG72}, which involves the stripping of the cold gas reservoir from the interstellar medium (ISM) of a galaxy due to pressures exerted by the hot, dense gas that permeates groups and clusters, quenches galaxies on timescales comparable to the crossing time ( or dynamical time), $\tau_{\mathrm{dyn}} \equiv R/V$, where $R$ and $V$ are the radius and velocity of the host dark matter halo, respectively.


Several environmental quenching studies have successfully constrained $\tau_{\mathrm{quench}}$. Notably, studies in the nearby Universe ($z\lesssim0.1$) constrain $\tau_{\mathrm{quench}}$ across a wide range of stellar masses in both galaxy groups and clusters \citep[e.g.,][]{Wetzel13, wheeler14, Fham15, Fham16, RodriguezWimberly19}. Among other findings, these studies indicate that satellite quenching timescales for galaxies with $\mstar \sim 10^{9.5-10.5}$ in both galaxy groups and clusters are relatively long ($4-8$ Gyr), stellar mass dependent, and generally consistent with the total cold gas (H{\scriptsize I}+H$_{2}$) depletion timescale, suggesting that starvation (as opposed to more rapid mechanisms such as RPS) is the dominant quenching mechanism at this epoch.

At higher redshifts ($z\sim1$), environmental quenching studies have predominantly focused on the most massive galaxy clusters ($\mhalo \sim 10^{14-15}~\msun$), where membership can be robustly assigned, and environmental processes are expected to be more pronounced \citep[e.g.,][]{Muzzin12, Stanford14, vdB14, Balogh21, Kim23}. At these redshifts $\tau_{\mathrm{quench}}$ for galaxies with $\mstar \gtrsim 10^{10.5}~\msun$ is $\sim1.5$ Gyr, again aligning with the total cold gas depletion timescale at this epoch \citep{Muzzin14, Balogh16, Foltz18, Baxter23}.

Conversely, studies of galaxy groups ($\mhalo \sim 10^{13-14}~\msun$) at $z\sim1$ report $\tau_{\mathrm{quench}}$ for galaxies with ($\mstar \gtrsim 10^{10.5}~\msun$) around $\sim 2.5$ Gyr \cite[e.g.,][]{Balogh16, Fossati17}. These results suggest that quenching in groups is due to the exhaustion of cold gas reservoirs in the absence of cosmological accretion, with modest feedback-driven outflows thought to play a secondary role \citep[see][]{McGee14, Balogh16}. However, it remains unclear how quenching proceeds for less massive ($\mstar \gtrsim 10^{9.5}~\msun$) group satellites at this epoch. Studies at $z \sim 0$ have shown that galaxies below this stellar mass threshold are more efficiently quenched by environment, making this an important regime to investigate at earlier times \citep{Peng10, Wetzel15, Fham15, Baxter21, Karunakaran23, Meng23}.

The objective of the Keck/DEIMOS survey presented in this investigation is to enhance the existing literature by assembling a mass-limited ($\mstar \gtrsim 10^{9.5}~\msun$) sample of spectroscopically-confirmed group satellite galaxies at $z\sim0.8$. This improves upon previous studies of low-mass galaxies in groups at this epoch, which have been limited to photometrically selected group members. This new data enables us to improve estimates of $\tau_{\mathrm{quench}}$ at $\mstar \gtrsim 10^{9.5}~\msun$, identify the dominant quenching pathway for galaxies in group environments at $z\sim0.8$, investigate the relative efficiency of environmental quenching in groups versus clusters, and constrain how quenching timescales of low-mass group satellites evolve with redshift.

This paper is structured as follows. In \S\ref{sec:KeckDEIMOS} we discuss how groups and group member candidates were selected and followed up with the DEep Imaging Multi-Object Spectrograph (DEIMOS) instrument using the Keck II telescope. We also describe the data reduction of the 1-D and 2-D spectra and present the resulting redshift catalog in \S\ref{sec:KeckDEIMOS}. In \S\ref{sec:Group-Membership} we describe our methodology for determining group membership, estimating physical properties via SED fitting, and separating group members in star-forming and quiescent sub-populations. In \S\ref{sec:quenching_model} we present satellite quenched fractions in bins of stellar mass, group-centric radius, and redshift, alongside a model that utilizes these measurements to constrain the satellite quenching timescales in groups at $z\sim0.8$. In \S\ref{sec:Results} we present our quenching timescale results. Finally, in \S\ref{sec:Discussion} we explore the implications for the dominant quenching mechanism in group environments at intermediate redshift and contextualize our results with respect to previous environmental quenching studies, before summarizing our findings in \S\ref{sec:Conclusion}. 

Throughout this study we adopt a Planck 2015 cosmology with $H_{0} = 67.7~{\rm km}~{\rm s}^{-1}~{\rm Mpc}^{-1}$ and $\Omega_{m}$ = 0.307 \citep{PlanckCollab16}, express distances in comoving units, magnitudes in the AB system (Oke and Gunn, 1983), and halo masses and sizes in terms of $\rtwo$, the radius within which the average density is $200$ times the critical density of the Universe at the redshift of the group.

\section{Data}
\label{sec:KeckDEIMOS}

\subsection{Keck/DEIMOS campaign overview}
\label{subsec:survey_overview}
This paper presents new data derived from a multi-cycle Keck/DEIMOS campaign, focused on obtaining high-resolution spectroscopy for suspected satellite galaxies in spectroscopically-confirmed galaxy groups in the Extended Groth Strip \citep[EGS,][]{Newman13}. The survey aimed to achieve a balanced representation of both star-forming and quiescent galaxies by selecting objects in the $K$-band ($K \lt 23.1$), with the objective of building a mass-limited sample down to $\mstar \sim 10^{9.5}~\msun$. Data were taken during the 2016A-2019A semesters, for a total of 9 nights.

\subsection{Group selection}
\label{subsec:group_selection}

We target $23$ galaxy groups identified from DEEP2 and DEEP3 spectroscopy \citep{Newman13, Cooper12, Zhou19} in the Extended Groth Strip (EGS). The targets are split between primary and secondary targets. The primary target sample consists of $11$ X-ray selected galaxy groups from the group catalog presented in \citet{Erfanianfar13}. These targets are a subsample of $52$ groups identified in a blind survey for X-ray extended sources from $800$ ks \textit{Chandra} coverage of the All-wavelength Extended Groth Strip International Survey \citep[AEGIS,][]{Davis07}. The optical counterparts of these X-ray sources were identified using CFHTLS-Wide3 imaging data, which is covered in $u^{*}, g^{'}, r^{'}, i^{'},$ and $z^{'}$ filters down to $i^{'} = 24.5$. Spectroscopic group membership was determined using DEEP2 and DEEP3 spectroscopy. However, photometric redshifts were used to confirm the overdensity of red galaxies within or near X-ray sources that lacked spectroscopic data, following the red sequence technique presented in \citet{Finoguenov10}. 

As detailed in \citet{Erfanianfar13}, each group is given a quality flag between 1 and 4, with one being the highest quality. Specifically, FLAG = 1 indicates that the group has a confident redshift assignment, significant X-ray emission, and good X-ray centering; FLAG = 2 indicates large uncertainties on the X-ray centering; FLAG = 3 indicates good centering but no spectroscopic confirmation; FLAG = 4 indicates large uncertainties on centering and uncertain redshifts due to lack of nearby spectroscopic objects and red galaxies; and Flag = 5 indicates that a redshift could not be identified.

10 out of the 11 group targets that comprise our primary sample have quality flags less than 4, between $2-13$ spectroscopically-confirmed group members, and are located in the redshift range $0.66 \lt z \lt 1.02$. Each group has a total estimated halo mass, $\mtwo$, which is determined using the X-ray luminosity-halo mass scaling relation from \citet{Leauthaud10}, assuming a standard evolution: $\mtwo E_{z} = f(L_{x}, E^{-1}_{z})$ where $E_{z} = (\Omega_{M}(1+z)^{3} + \Omega_{\Lambda})^{1/2}$. The primary group sample has total halo masses that range from $\mtwo \sim 2-7 \times 10^{13}~\msun$.

To enhance the efficiency the survey, our observing masks were configured to include an additional $12$ optically-selected groups in the redshift range $0.53 \lt z \lt 1.03$. These groups were initially identified in the publicly-available catalog of galaxy groups by \citet{Gerke12}, which applied the Voronoi-Delauny Method (VDM) group finder to spectroscopic data from the fourth data release of the Deep Extragalactic Evolutionary Probe 2 (DEEP2) spectroscopic survey \citep{Davis03, Newman13}. The main difference between the primary and secondary target samples is that the latter lacks X-ray observations. As discussed in \S\ref{subsec:Group-Membership}, these secondary systems contribute minimally to our analysis, with only two of the twelve groups having dynamically-estimated halo masses typical of galaxy groups ($10^{13-14}~\msun$). Consequently, we exclude these secondary targets from our analysis, focusing exclusively on groups with confirmed extended X-ray emission, a strong indicator of virialization. Thus, this selection criterion ensures that our analysis is limited to galaxy groups likely to be virialized, with relatively robust halo mass estimates.

\subsection{Spectroscopic target selection}
\label{subsec:target_selection}

The targets for spectroscopic follow-up are sourced from NEWFIRM Medium-Band Survey \citep[NMBS,][]{Whitaker11} photometry.
Specifically, we utilize version 5.1 of the deblended catalog with sources in the All-wavelength Extended Groth Strip International Survey \citep[AEGIS,][]{Davis07} field\footnote{\url{http://www.astro.yale.edu/nmbs/Data_Products.html}}. This deep, multi-wavelength dataset, which features $ugrizJ_{1}J_{2}J_{3}H_{1}H_{2}K$ photometry, complemented by space-based imaging from \textit{HST}/ACS, \textit{Spitzer}, and GALEX, provides accurate photometric redshifts ($\sigma_{z}/(1+z)=0.017$) for efficient selection of group member candidates.

In addition, we leverage existing spectroscopy from DEEP2/DEEP3 to construct a cross-matched catalog, constructed by enforcing a maximum separation of 1 arcsecond between objects from the photometric (NMBS) and spectroscopic (DEEP2/DEEP3) catalogs. This merged catalog, which serves as the primary source for our science targets, allow us to easily deprioritize galaxies with existing DEEP2/DEEP3 spectroscopy. Additionally, we include overlapping SDSS photometry to identify guide and alignment stars in the EGS. These stars, restricted to objects below the SDSS magnitude limit of $R\lesssim22$, comprise our \textquote{guide catalog}. We use the guide catalog as a reference to align the primary science catalog by measuring and applying the relative astrometric offset.

With the aim of constructing a stellar-mass limited sample of $\gtrsim 10^{9.5}\msun$ at $z \sim 1$, we adopt a $K$-band selection limit of $K \lt 23.1$ (or $R\lesssim25.5\footnote{This limit is derived from examining the $R$-band distribution of galaxies that satisfy the $K$-band selection limit.}$). Notably, this depth exceeds that of the DEEP2 and DEEP3 surveys \citep[$R_{\mathrm{lim}} = 24.1$;][]{Newman13, Cooper12} by more than 1 magnitude. Utilizing this $K$-band selection criterion, we identify group candidates as galaxies with comoving separations of less than 1 Mpc from a group center and rest-frame redshift separations $\Delta z/(1+z) \lt 0.15$. 
In total, approximately 1500 satellite candidates around 23 spectroscopically-confirmed groups in the EGS are targeted in this survey.

\begin{figure}
\centering
\hspace*{-0.25in}
\includegraphics[width=3.5in]{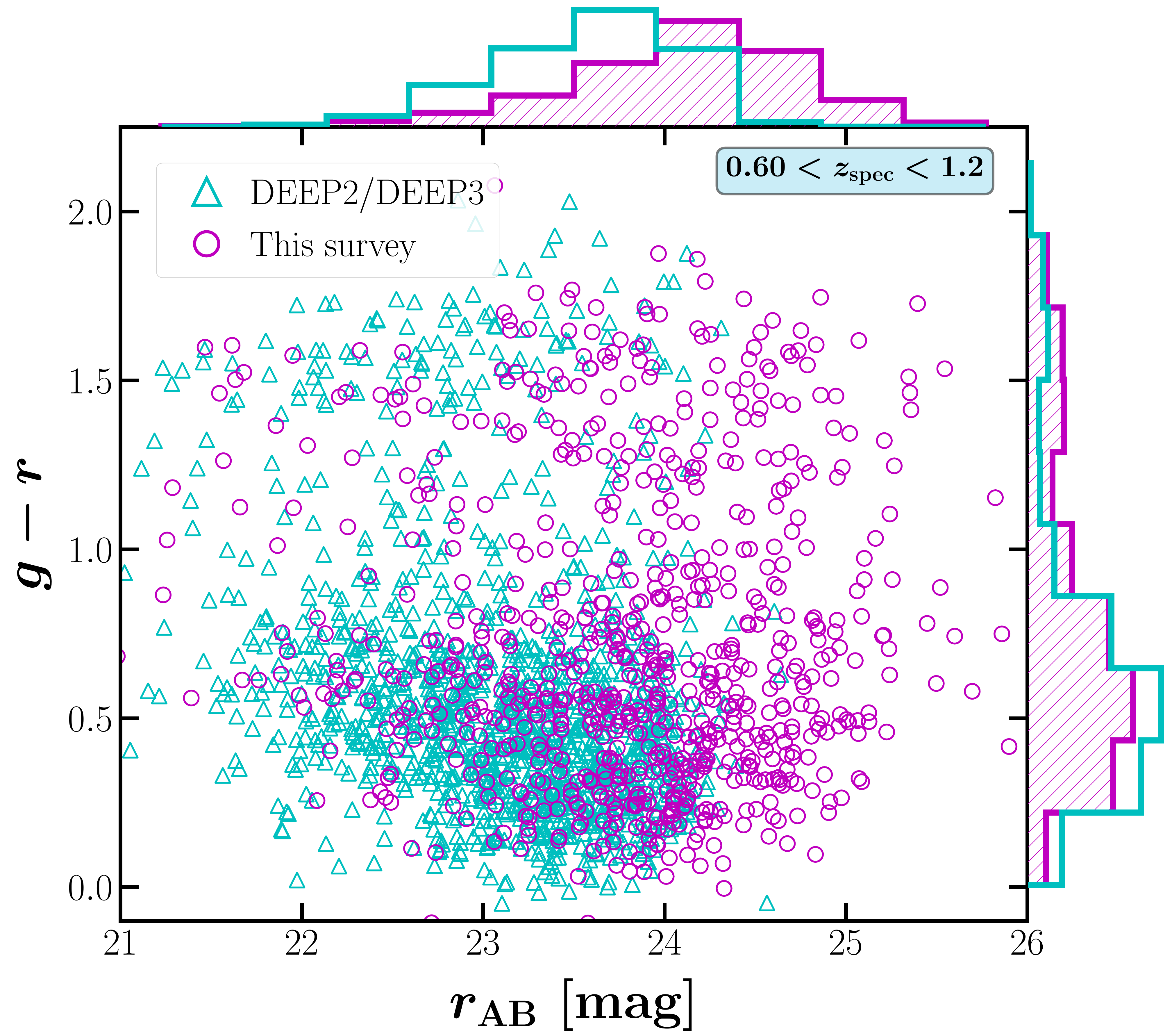}
\caption{The color-magnitude diagram shows the $g-r$ distribution for galaxies in EGS from DEEP2/DEEP3 and our Keck/DEIMOS survey. The data shown is restricted to galaxies with high-quality spectroscopic redshifts within the range $0.60 < z < 1.2$, aligning with the redshift range of our EGS group sample. Our observations enhance the existing data by adding high-quality spectroscopic redshifts beyond the $R_{\mathrm{AB}} = 24.1$ limiting magnitude of the DEEP2/DEEP3 survey.}
\label{fig:fig1}
\end{figure}

\begin{table*}
\centering
\caption{Galaxy redshift catalog}
\setlength{\tabcolsep}{9pt} 
\begin{tabularx}{\textwidth}{ccccccccc}
\hline
\hline
Object ID$^a$ & Mask$^b$ & Slit$^c$ & MJD$^d$ & RA$^e$ & DEC$^f$ & $z_{\mathrm{obs}}$$^g$ & $z_{\mathrm{helio}}$$^h$ & $z_{\mathrm{quality}}$$^i$ \\
\midrule
1961 & m19A16 &  0   & 58628.4102 & 214.44531 & 52.40640 &             0.32785 &               0.32790 &                       3 \\
      2152 & m19A16 &  1   & 58628.4102 & 214.37834 & 52.41012 &             0.97771 &               0.97775 &                       3 \\
      2238 & m19A16 &  4   & 58628.4102 & 214.42816 & 52.41250 &             0.61744 &               0.61748 &                       2 \\
      2327 & m19A16 &  5   & 58628.4102 & 214.48625 & 52.41427 &             0.31458 &               0.31462 &                       2 \\
      2609 & m19A16 &  6   & 58628.4102 & 214.43631 & 52.41878 &             0.42200 &               0.42204 &                       4 \\
      2616 & m19A16 &  7   & 58628.4102 & 214.35102 & 52.41954 &             0.81242 &               0.81246 &                       4 \\
      2648 & m19A16 &  8   & 58628.4102 & 214.46630 & 52.41904 &             0.44642 &               0.44647 &                       2 \\
      2786 & m19A16 &  10  & 58628.4102 & 214.42449 & 52.42131 &             0.54177 &               0.54181 &                       4 \\
\hline
\hline
\label{table:redshift_catalog}
\label{table:1}
\end{tabularx}
\footnotesize
\begin{flushleft}
$^{a}$Unique object identification number from the Newfirm Medium Band Catalog. \\
$^{b}$Name of slit mask on which the object was observed. \\
$^{c}$Slit number on the corresponding mask. \\
$^{d}$Modified Julian date of the observation. \\
$^{e}$Right ascension in degrees. \\
$^{f}$Declination in degrees. \\
$^{g}$Measured spectroscopic redshift from DEIMOS spectrum. \\
$^{h}$Heliocentric-corrected redshift \\
$^{i}$Redshift quality flag ($-2 =$ poor quality spectrum; $-1 = $ star; $1 =$ very low S/N; $2 =$ uncertain redshift measurement; $3, 4 =$ high quality redshift measurement) \\
\tablecomments{Table 1 is published in its entirety in the machine-readable format. A portion is shown here for guidance regarding its form and content.}
\end{flushleft}
\end{table*}

\subsection{Observations and data reduction}
\label{subsec:observations}


The spectroscopic follow-up observations were conducted using the DEep Imaging Multi-Object Spectrograph (DEIMOS) mounted on the 10m Keck II telescope. 
DEIMOS was chosen for its capability to measure spectroscopic redshifts at $z \sim 1$, leveraging the $\oii$ doublet and Ca H $\&$ K absorption features to probe both star-forming and quiescent galaxies. 
The instrument was configured with the 1200 lines mm$^{-1}$ grating blazed at $7500$\AA~and centered at $\sim 7750$\AA~(with the OG550 order-blocking filter), providing high resolution for precise spectroscopic redshifts ($\sigma_{z} \sim 30$ $\kms$), crucial for confirming group membership.

A total of 17 slitmasks were manufactured using the DSIMULATOR software\footnote{\url{https://www2.keck.hawaii.edu/inst/deimos/dsim.html}}, a mask design and optimization tool. 
The masks were designed with efficiency in mind, allowing $\sim130-160$ targets per mask, with $1\arcsec$ slit widths and minimum slit lengths of $4\arcsec$. 
%
%
Furthermore, the slit gaps were designed to have $0.5\arcsec$ separations, and the slits were tilted $5$ degrees relative to the mask position angle to increase the wavelength sampling of the sky region to aid in sky subtraction. 
Before being observed, each target was given a priority for observation categorized as either extremely faint ($R \gt 25.5$), moderately faint ($24 \lt R \lt 25.5$), or relatively bright group candidates ($R \lt 24.3$).
Masks were designed with this priority in mind to preferentially target brighter sources for nights with unfavorable observing conditions. 

Over the course of the survey, observations were acquired with a typical seeing of $0.68\arcsec$, determined by analyzing the median full width at half maximum (FWHM) of alignment stars across the spectral range of $7750$\AA-$9100$\AA. 
In general, we used total exposure times of $\gtrsim 6$ hours, which were selected to effectively capture spectra for low-mass ($\sim 10^{9.5}~\msun$), quiescent group member candidates. 
However, for nights with unfavorable observing conditions, either due to high humidity and/or poor transparency, we prioritized masks with preferentially brighter targets and adopted relatively shorter total exposure times of $\sim2$ hours.


All data acquired from the survey were reduced using the spec2D DEEP2/DEEP3 DEIMOS data reduction pipeline \citep{Cooper12, Newman13}. 
Redshifts were determined using the spec1d automated redshift pipeline, supplemented with visual redshift verification using zspec. 
The DEEP2/DEEP3 classification scheme was used to assign quality codes  ($z_{\mathrm{quality}})$ to each visually inspected redshift.
These codes are as follows: $z_{\mathrm{quality}} = -2$ (poor quality spectrum/effectively unobserved); $z_{\mathrm{quality}} = -1$ (star); $z_{\mathrm{quality}} = 1$ (likely an extended object, but very low S/N); $z_{\mathrm{quality}} = 2$ (low S/N and/or compromised data leading to uncertain redshift measurement); $z_{\mathrm{quality}} = 3$ (reliable redshift with probability of accuracy $\geq 95\%$); $z_{\mathrm{quality}} = 4$ (reliable redshift with probability of accuracy $\geq 99\%$).

\subsection{Galaxy redshift catalog}
\label{subsec:Data Reduction}

We provide the spectroscopic data obtained from our Keck/DEIMOS observations in Table~\ref{table:1} (the complete version can be found in the digital edition of the Journal). For the purpose of this investigation, we consider only galaxies with secure high-quality redshifts ($z_{\mathrm{quality}} \geq 3$) for which there are $1313$ objects, including $1040$ without previous spectroscopic redshifts. Although our sample is relatively small compared to DEEP2 and DEEP3, the extension to fainter magnitudes (as illustrated in Fig.~\ref{fig:fig1}) enhances the utility of each obtained spectroscopic redshift. 

The final catalog comprises $7058$ unique redshifts, sourced from this survey ($1313$), DEEP2/DEEP3 ($5702$), and overlapping SDSS coverage ($43$).

\section{Observed Group Sample}
\label{sec:Group-Membership}

\begin{table*}
\centering
\caption{Properties of 11 X-ray Selected Groups}
\setlength{\tabcolsep}{10pt} 
\begin{tabularx}{\textwidth}{ccccccccc}
\hline
\hline
\multirow{2}{*}{Group ID$^{a}$} &   RA$^{b}$ &  DEC$^{c}$ &   \multirow{2}{*}{$z$$^{d}$} &   $\sigma$$^{e}$ &  $R_{200}$$^{f}$ &  $M_{200}$$^{g}$  &{$N_{\mathrm{mem}}$$^{h}$} & Flag$^{i}$ \\
     
     & [\rm deg] & [\rm deg] & & [km s$^{-1}$] & $[\rm ckpc]$ & [$10^{13}~{\rm M}_{\odot}$] & [$< R_{200}$] &\\
\midrule
J1416.4+5227 & 214.12029 & 52.45258 & 0.837 &                  386.0 &   1123.170 &       6.26 &       13 (7) & 1.0 \\
J1416.6+5228 & 214.16351 & 52.48248 & 0.812 &                  307.0 &    928.283 &       3.17 &              8 (4) & 3.0  \\
J1417.0+5226 & 214.25454 & 52.44760 & 1.023 &                  340.0 &   1004.364 &       3.85 &                 7 (4) & 1.0 \\
J1417.5+5232 & 214.39395 & 52.53531 & 0.985 &                  399.0 &   1172.309 &       6.35 &               14 (5) & 2.0  \\
J1417.5+5238 & 214.38802 & 52.63349 & 0.717 &                  358.0 &   1070.654 &       5.32 &               23 (13) & 2.0  \\
J1417.7+5228 & 214.44257 & 52.46947 & 0.995 &                  372.0 &   1082.625 &       5.12 &                12 (4) & 1.0  \\
J1417.9+5225 & 214.48623 & 52.42069 & 0.995 &                  347.0 &    984.205 &       4.13 &                20 (7) & 3.0 \\
J1417.9+5226 & 214.48705 & 52.43759 & 0.684 &                  301.0 &    883.415 &       3.22 &                 3 (2) & 4.0 \\
J1417.9+5231 & 214.47931 & 52.52391 & 0.661 &                  308.0 &    930.701 &       3.49 &                3 (2) & 1.0  \\
J1417.9+5235 & 214.47645 & 52.58274 & 0.683 &                  346.0 &   1029.427 &       4.92 &              27 (12) & 1.0 \\
J1418.8+5248 & 214.68967 & 52.80612 & 0.741 &                  274.0 &    785.177 &       2.36 &                5 (4) & 1.0  \\

\hline
\hline
\label{table:groups}
\label{table:2}
\end{tabularx}
\footnotesize
\begin{flushleft}
$^{a}$Group identification number. \\
$^{b}$Right Ascension in degrees, measured from the location of peak X-ray emission. \\
$^{c}$Declination in degrees, measured from the location of peak X-ray emission. \\
$^{d}$Redshift of peak X-ray luminosity. \\
$^{e}$X-ray-derived velocity dispersion.  \\
$^{f}$$\rtwo$ derived from X-ray luminosity.\\
$^{g}$$\mtwo$ derived from X-ray luminosity. \\
$^{h}$Number of spectroscopic group members found in this study, with those originally found in \citet{Erfanianfar13} listed in parentheses
$^{i}$ Group identification flag ( 1 = confident redshift assignment, significant X-ray emission, and good centering; 2 = centering has a large uncertainty; 3 = no spectroscopic confirmation but good centering; 4 = unlikely redshifts due to the lack of spectroscopic objects and red galaxies and also a large uncertainty in centering; 5 = could not identify any redshift.
\end{flushleft}
\end{table*}

\subsection{Group membership}
\label{subsec:Group-Membership}

Galaxies are classified as group members based on two selection criteria:

\begin{enumerate}
    \item The rest-frame line-of-sight velocity difference between a galaxy and a nearby group is limited to $|\Delta v| < 1000$ $\kms$, which is approximately three times the average velocity dispersion of the group sample.  
    
    \item The projected separation relative to the location of the peak X-ray emission is less than $R_{200}$ of the group.
\end{enumerate}

Applying these membership selection criteria leads to the identification of 134 spectroscopically-confirmed group members across the 11 X-ray-selected groups. We subsequently refine group membership by measuring separations relative to the location of the luminosity-weighted centroid of the group members instead of the location of peak X-ray emission. This adjustment serves two purposes: to facilitate a more accurate comparison with mock galaxy groups from $N$-body simulations and to account for the fact that only six of the eleven X-ray selected groups possess well-defined group centers based on the X-ray emission. We find that the separation between the location of the peak X-ray emission and the luminosity-weighted centroid ranges from $0.83\arcsec$ to $25.8\arcsec$, corresponding to $14$ kpc to $338$ kpc, with a median separation of $13.4\arcsec$ or $201$ kpc. Given that the typical absolute astrometric accuracy of \textit{Chandra} is approximately $0.6\arcsec$, these offsets are unlikely to be caused by astrometric errors. Instead, they may reflect physical properties of the X-ray emission, such as asymmetric morphologies or substructures within the groups. This is consistent with the findings of \citet{Erfanianfar13}, who noted that some X-ray sources exhibit secondary peaks or irregular emission, which could shift the X-ray peak relative to the luminosity-weighted galaxy centroid. Despite these spatial offsets, the membership of the groups remains remarkably stable, with the total number changing by only one member, from 134 to 135. This is unsurprising, given that the typical virial radius is 4-5 times larger than even the maximum observed offset between the X-ray emission center and the luminosity-weighted centroid. In column 8 of Table~\ref{table:2}, we compare the number of satellites confirmed from the membership analysis employed in this study with those previously confirmed in \citet{Erfanianfar13}. In nearly all cases, we reaffirm membership, with only four exceptions identified as interlopers.

\begin{figure*}
\centering
\hspace*{-0.1in}
\includegraphics[width=7.0in]{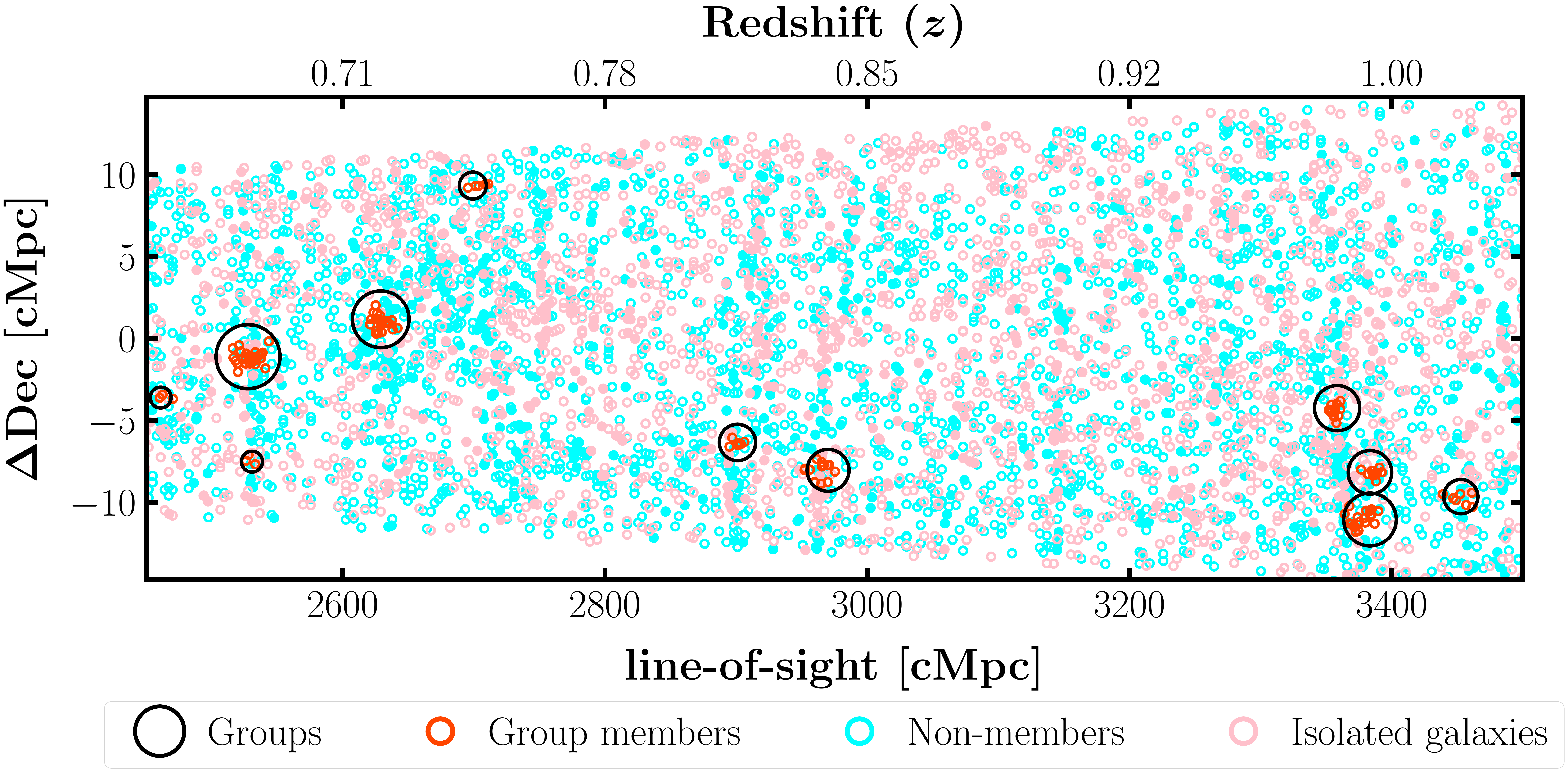}
\caption{The redshift-space distribution of galaxies in the EGS field is shown. The vertical axis represents the transverse distance along the declination direction on the sky. The black circle marks the location of our groups at the luminosity-weighted centroid, with the circle size indicating the group's richness. Orange circles represent the spectroscopically confirmed group members. Cyan and pink circles represent non-group members, with closed and open circles distinguishing between galaxies with and without spectroscopic redshifts, respectively. The pink circles specifically highlight the subset of non-member galaxies that constitute our control sample of isolated galaxies.}
\label{fig:fig2}
\end{figure*}

In addition to the aforementioned membership selection technique, we also test two alternative techniques.
The first, which includes both the primary and secondary groups, follows the membership selection technique presented in Section 5 of \cite{Erfanianfar13}.
This technique relies on assuming an initial velocity dispersion for each group, with our choice of $\sigma=500~\kms$ matching that assumed by \cite{Erfanianfar13}.
This initial velocity dispersion is then used to calculate both a maximum redshift threshold and a projected spatial separation threshold for determining group membership.
Initial group members are selected based on these thresholds and then refined iteratively using the \textquote{gapper} method to compute the observed group velocity dispersion.
The gapper method is an iterative technique that calculates the velocity dispersion by analyzing the gaps between the measured line-of-sight velocities within a given group, after sorting the velocities in ascending order. It has been shown to be an accurate estimator, particularly in the limit of small sample sizes \cite{Beers90}.

The gapper method derived velocity dispersion is iteratively used to reassess group membership until stability is achieved, which we reach after two iterations. 
The final velocity dispersion is subsequently used to compute the rest-frame and intrinsic velocity dispersion for each group, where the latter is achieved by removing the effect of measurement errors of component galaxies from the rest-frame velocity dispersion. 
The intrinsic velocity is then used to compute $\rtwo$ and subsequently $\mtwo$, under the assumption of virialization.
From this exercise we find that the membership of the primary X-ray-selected group sample is relatively unchanged, whereas only five of the twelve secondary groups yield halo masses in the conventional group halo mass range of $\mtwo = 10^{13-14}~\msun$.
Notably, three of these five groups are located at lower redshifts ($z=0.53-0.58$) compared to the rest of the sample and are thus excluded from this analysis.
We ultimately find that the primary measurable from this investigation, the satellite quenched fraction, is robust to the inclusion and exclusion of these two additional groups. Therefore, for the sake of simplicity, we continue using the membership selection technique based on the line-of-sight velocity and projected spatial separation from the luminosity weighted centers of the X-ray selected groups.

Second, we investigate including photometric redshift members to increase the number of low-mass quiescent galaxies in our sample.
While maintaining the originally defined projected separation threshold, we test two rest-frame line-of-sight velocity thresholds for galaxies lacking spectroscopic redshifts.
The first conservative approach defines photometric members as galaxies with $|\Delta v| < 1000$ $\kms$, resulting in an increase in group membership from 135 to 149.
The second, less conservative approach defines photometric members as galaxies with $|\Delta v| < 2500$ $\kms$, leading to a rise in group membership from 135 to 167.
Our analysis reveals minimal impact on the quiescent fraction at both high and low stellar masses when including photometrically selected members.
Moreover, we find that the differences in quiescent fractions resulting from the inclusion of photometric members are substantially smaller than the uncertainties associated with the quenched fraction derived solely from spectroscopically confirmed members.
Consequently, to ensure membership purity and minimize contamination from interlopers, we opt to retain the more reliable spectroscopic selection criteria for the remainder of the analysis.

\subsection{Control sample}
\label{subsubsec:control-sample}

We utilize the NMBS catalog to establish a control sample of \textquote{isolated} galaxies in the EGS. 
This catalog contains approximately $22500$ $K$-band selected galaxies in the AEGIS field with accurate photometric redshifts and rest-frame colors, as well as stellar masses and star-formation rates derived using the Fitting and Analysis of Spectral Templates \citep[FAST][]{Kriek09} spectral energy distribution (SED) fitting code.
From this initial sample, we limit to galaxies in the redshift range $0.6 \lt z \lt 1.2$, which yields $\sim9100$ galaxies, of which $\sim23~\%$ have spectroscopic redshifts. 
From this sample we define isolation as being sufficiently far away, both in terms of line-of-sight velocity and projected separation, from any massive ($\gt 10^{10.75}~\msun$) galaxy.  
This specific threshold is selected based on the stellar mass distribution of the two most massive group members for each group, with $\mstar \sim 10^{10.75}\msun$ being the minimum of the distribution.
We identify $596$ massive galaxies and establish the isolation criteria as follows: 

\begin{enumerate}
    \item A galaxy must have a comoving separation greater than $2$ cMpc (i.e., twice the typical $\rtwo$ of the observed group sample) from any massive galaxy, excluding itself.

    \item The rest-frame line-of-sight velocity difference between a galaxy and any massive galaxy must be greater than $|\Delta v| \gt 5000~\kms$ (five times the line-of-sight velocity cut used to determine group membership), again excluding itself. This threshold is reduced to $|\Delta v| \gt 1500~\kms$ for galaxies with spectroscopic redshifts. 
\end{enumerate}

Applying these criteria we identify approximately $3600$ isolated galaxies in the redshift range $0.6 < z < 1.2$.
Figure~\ref{fig:fig2} shows the redshift-space distribution for groups (unfilled black circles), group members (unfilled orange circles), non-group members (cyan circles), and isolated galaxies (pink circles).
Open and closed circles indicate galaxies with and without a spectroscopic redshift, respectively, for non-group members and isolated galaxies. 
This figure illustrates the locations of the galaxy groups and the control sample of isolated galaxies in relation to the large-scale structure, as traced by the observed galaxy population.
As detailed in \S\ref{subsec:sim_class}, we use the control sample to estimate the fraction of isolated quenched galaxies as a function of redshift and stellar mass. 
This is a crucial measurement used to constrain the contribution to the observed satellite quenched fractions from satellites that likely quenched independent of the group environment.

\subsection{Physical properties from SED fitting}
\label{subsec:stellar_mass}

We determine the physical properties of the confirmed group members using the Bayesian Analysis of Galaxies for Physical Inference and Parameter Estimation (BAGPIPES) SED fitting code \citep{Carnall18}. We fit all available NMBS photometry, which includes coverage in the UV (\textit{GALEX}), visible and NIR (Canada-France-Hawaii Telescope and Subaru Telescope), and mid-IR (\textit{Spitzer}/IRAC). 
During the fitting process we assume an exponentially-declining star formation history (SFH) given by SFR $= \mathrm{exp}^{-t/\tau}$ and fix each galaxy to its spectroscopic redshift. The following parameter ranges are utilized for fitting: time since SFH began $0.1~\mathrm{Gyr} < t < 15~\mathrm{Gyr}$; timescale of declination $0.3~\mathrm{Gyr} < \tau < 10~\mathrm{Gyr}$; formation mass $1.0 < \mathrm{log}_{10}(\mstar/\msun) < 15$; metallicity $0 < \mathrm{Z}/\mathrm{Z_{\odot}} < 2.5$; and reddening $0 < A_{\rm{v}} < 4.0$ (using a \citet{Calzetti00} dust law). Furthermore, we adopt the default stellar population models, namely the 2016 version of the \citet{BruzualCharlot03} models.

We compare the stellar masses derived using BAGPIPES with those listed in the deblended NMBS catalog, which were obtained using FAST. While the two sets of values are generally consistent, we note a moderate offset of $\sim 0.1$ dex, with our updated stellar masses being larger than those obtained using FAST. This discrepancy is likely attributable to differences in SED fitting methodologies, model assumptions, and a minor influence of variations between our spectroscopic redshifts and the photometric redshifts employed by FAST for stellar mass estimation.

\subsection{Galaxy classification}
\label{subsec:classification}
We classify group members as star-forming or quiescent based on rest-frame UVJ colors, using the color-color cuts defined by \citet{Whitaker11} and shown in Equation~\ref{uvj_eqn}.

\begin{equation}
\begin{split}
&(U-V) > 1.3 \quad \cap \\
&(V-J) < 1.6 \quad \cap\\
&(U-V) > 0.88 \times (V-J) + 0.59 \; .
\end{split}
\label{uvj_eqn}
\end{equation}

Although the NMBS catalog provides rest-frame UVJ colors, these values are based on photometric redshifts.
Given the modest scatter between photometric and spectroscopic redshifts of group members ($\sigma_{z} \sim 0.06$), we remeasure the rest-frame UVJ colors using the spectroscopic redshifts from our Keck/DEIMOS observations.
To maintain consistency with the NMBS catalog, we use the EAZY code \citep{Brammer08}, applying the same configuration file as used in the catalog’s construction. 
Using Equation~\ref{uvj_eqn}, we identify 62 quenched galaxies and 73 star-forming galaxies. 
For comparison, these numbers are 61 and 74, respectively, when using the rest-frame UVJ colors provided in the NMBS catalog.

Additionally, we classify group member galaxies based on the median specific star formation rates (sSFR) estimated using BAGPIPES. 
We apply a visually selected threshold of $\text{sSFR} = 10^{-10.35}~\text{yr}^{-1}$, which effectively separates the star-forming and quiescent populations, and coincidentally aligns with the redshift-dependent sSFR threshold of $0.3 \times t_{\rm{Hubble}}^{-1}$ from \citet{Franx08}.
This criterion yields 52 quenched and 83 star-forming galaxies. Upon comparison, the two classification methods demonstrate significant agreement, with $\gt 90\%$ of the galaxies having consistent classifications. 
However to ensure consistency across our analysis we adopt the rest-frame UVJ color-color classification. 
This choice is influenced by two factors. 
First, it is essential to maintain a consistent classification system for both group and control samples. 
Second, a considerable portion of the control sample lacks reliable sSFR estimates.


%

\begin{figure}
\centering
\hspace*{-0.25in}
\includegraphics[width=3.5in]{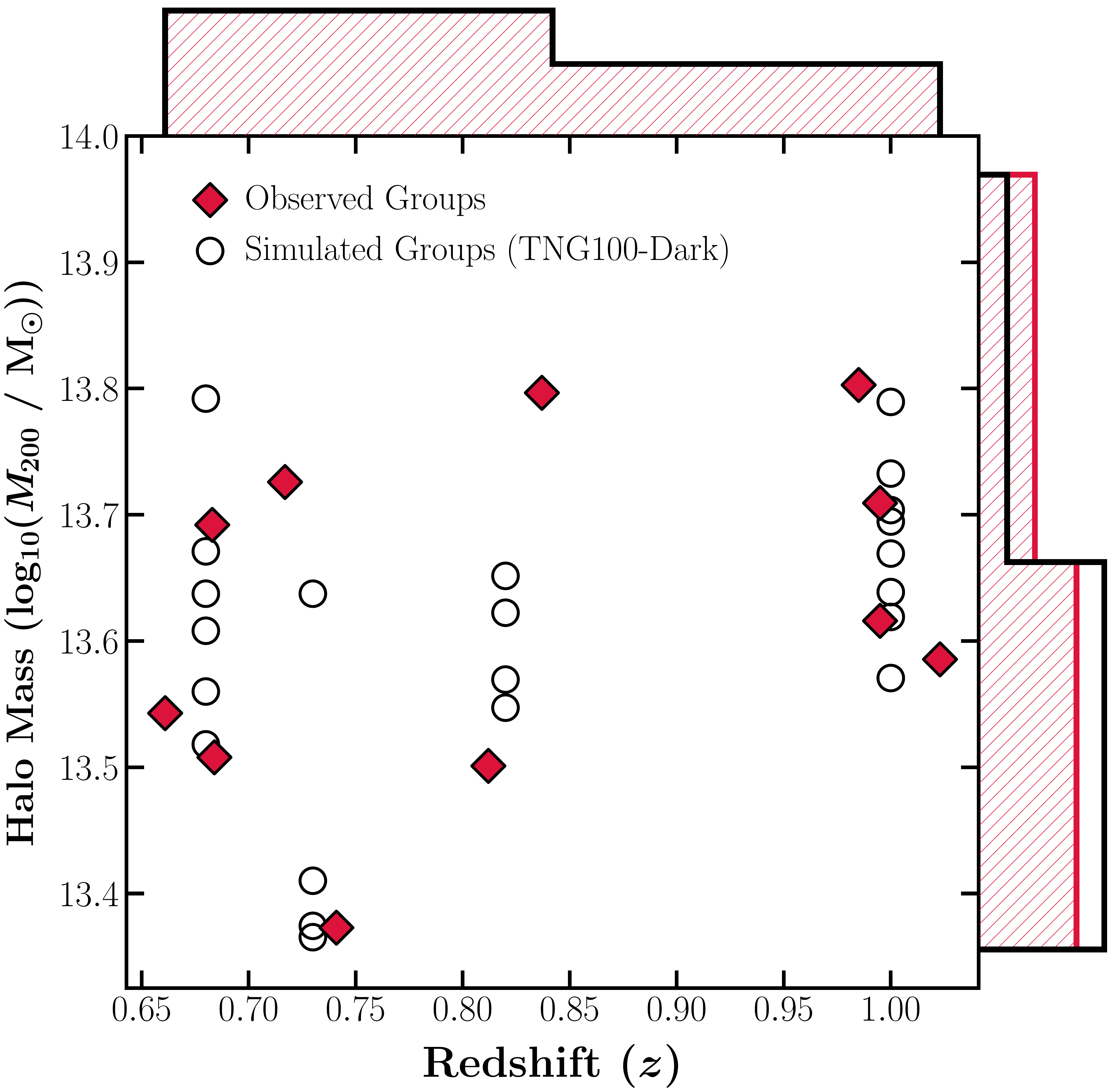}
\caption{$\mtwo$ versus $z$ for the observed and simulated group samples. The open circles (filled diamonds) represent the TNG100-Dark (EGS) groups.}
\label{fig:fig3}
\end{figure}

\begin{figure}
\centering
\hspace*{-0.25in}
\includegraphics[width=3.5in]{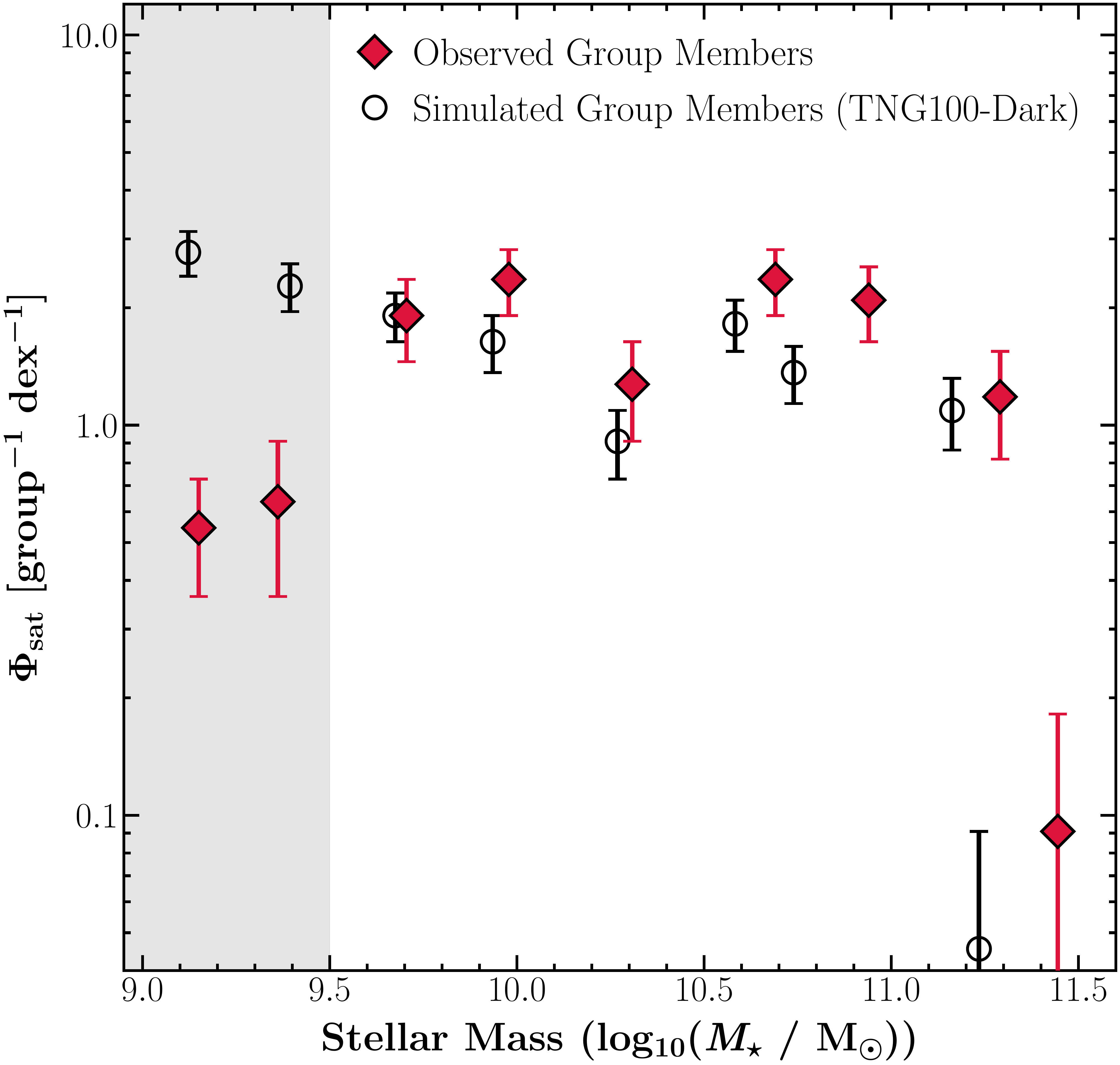}
\caption{Comparison of the satellite stellar mass function between observed (red diamonds) and simulated (open circles) groups. The observed sample exhibits a slight extension towards higher masses compared to the simulated counterpart, yet both distributions demonstrate relative agreement down to approximately $10^{9.5}~\msun$. Our analysis focuses solely on observed and simulated group members above $10^{9.5}~\msun$, as highlighted by the shaded gray band.}
\label{fig:fig4}
\end{figure}

\section{Simulated Group Sample}
\label{subsec:TNG100-Dark}

\subsection{Simulated group selection}
\label{subsec:simulated_groups}

We use the TNG100-1-Dark simulation from the IllustrisTNG project\footnote{https://www.tng-project.org} \citep[TNG;][]{Nelson18, Niaman18, Springel18, Pillepich18, Marinacci18} to track the distribution of infall times (and thus quenched fraction given an assumed quenching timescale) for a simulated group population selected to closely approximate the redshift and halo mass distribution of the observed group sample. The TNG100-1-Dark simulation (hereafter referred to as TNG100-Dark) is a state-of-the-art, high-resolution cosmological model that simulates the formation and evolution of galaxies and large-scale structure in the Universe. TNG100-Dark covers a cubic region with a side length of $75~\mathrm{cMpc}/h$ and achieves a dark matter mass resolution of $6 \times 10^{6}~\msun/h$. This combination of volume and resolution enables us to reliably investigate analogs of our observed group population.

The simulated groups are derived from the group catalogs and sublink merger trees of the TNG100-Dark simulation, which encompasses a total of 100 snapshots from $z=20.05$ to $z=0$. Our analysis focuses on a subset of 4 snapshots at $z_{\mathrm{snap}}=1.0,~0.82,~0.73$, and $0.68$, chosen to mirror the redshift distribution of the observed group sample. Each snapshot contains a distinct group catalog comprising both friends-of-friends (FoF) groups \citep{Davis85} and \textsc{Subfind} objects \citep{Springel01, Dolag09}. The FoF catalog includes a GroupFirstSub column, which identifies the primary/central as the most massive subhalo  within each FoF group. 

For each snapshot we restrict groups to have halo masses ranging from $10^{13.3}~\msun$ to $10^{13.8}~\msun$. We then utilize the TNG100-Dark Sublink merger trees to track the \textsc{Subfind} IDs of these groups from $z=0.68$ to $z=1.0$, ensuring the identification of unique groups across the $4$ snapshots without any progenitor-descendant relation. In other words, a group selected in a higher redshift bin is not a progenitor of a group in a lower redshift bin. After constructing the sample of unique groups, we further limit the groups in each snapshot to have halo masses approximately within the range of the observed groups at a given redshift. As shown in Fig.~\ref{fig:fig3}, our simulated group sample comprises a total of $22$ unique groups from snapshots spanning from $z=1.0$ to $z=0.68$. This simulated sample has a median redshift of $\tilde{z}=0.82$ and a median halo mass of $\tilde{M}_{\rm{200}}= 10^{13.63}~\msun$, consistent with the median redshift ($\tilde{z}=0.81$) and a median halo mass of ($\tilde{M}_{\rm{200}}=10^{13.62}~\msun$) of the observed group sample.

\subsection{Simulated group membership}
\label{subsec:simulated_group_membership}

We define simulated group members as all subhalos (including the central/primary, defined as the most massive subhalo in a given group) that satisfy the condition $d_{\rm{host}}(z_{\rm{obs}}) \leq R_{200}$, where $d_{\rm{host}}(z_{\rm{obs}})$ represents the three-dimensional comoving separation between a subhalo and the group center as measured at the observed redshift. Once the satellite population is established, we utilize the TNG100-Dark sublink merger trees to track subhalo properties such as position, halo mass, and $R_{200}$ along the main progenitor branch from $z=20.05$ to $z_{\rm obs}$.

Since TNG100-Dark is a dark-matter-only simulation we estimate stellar masses using the empirical stellar mass-to-halo mass (SMHM) relation proposed by \citet{Behroozi19}. This relation allows us to assign stellar masses to all subhalos at their respective snapshots using their peak halo mass at a given $z_{\mathrm{snap}}$. For this procedure we use the SMHM fit parameters and their uncertainties listed in rows 2 and 3 of Table J1 in \cite{Behroozi19} to assign stellar masses to centrals and satellites, respectively. Specifically, we use the spread in the SMHM relation, as quantified by the $68\%$ confidence interval from the model posterior space, to assign stellar masses via Gaussian sampling.  

In Fig.~\ref{fig:fig4} we compare the stellar mass distribution of the simulated group sample with the observed sample in bins of size $0.25$ dex. The mass-resolution of the TNG100-Dark simulation enables the reliable detection of low-mass satellites down to $\sim10^{9.0}~\msun$. However, the observed group sample shows signs of incompleteness below $\mstar \lesssim 10^{9.5}~\msun$. For the purpose of our analysis we limit both the observed and simulated samples to group members with $\mstar \gt 10^{9.5}~\msun$, which yields $203$ satellites across the $22$ simulated groups and $122$ satellites across $11$ EGS groups.

\section{Quenching model}
\label{sec:quenching_model}

We utilize an infall-based modeling framework akin to that utilized in \citep{Wetzel13, wheeler14, Fham15, Baxter22} to estimate the satellite quenching timescale for our observed group sample. Unlike quenching timescales derived from star formation histories, which measure the time between peak star formation and when the star formation rate falls below a specific threshold \citep[e.g.,][]{Carnall19}, our approach focuses on the timescale over which a galaxy quenches after it has become a satellite of a group or cluster. This approach leverages the assembly histories of group populations from $N$-body simulations to determine how long a subhalo of a given mass has been a group member. This information allows us to construct an infall-based quenching model that: i) probabilistically determines the fraction of galaxies that quench due to secular processes prior to infall, based on observed measurements of the isolated quenched fraction (see \S\ref{subsubsec:control-sample}); and ii) constrains the time required for galaxies to quench due to environmental mechanisms after infall. Specifically, the model generates satellite quenched fractions based on assumed satellite quenching timescales, and we use a Bayesian approach to identify the timescales that are most consistent with the observed satellite quenched fraction trends.

\begin{figure}
\centering
\hspace*{-0.25in}
\includegraphics[width=3.5in]{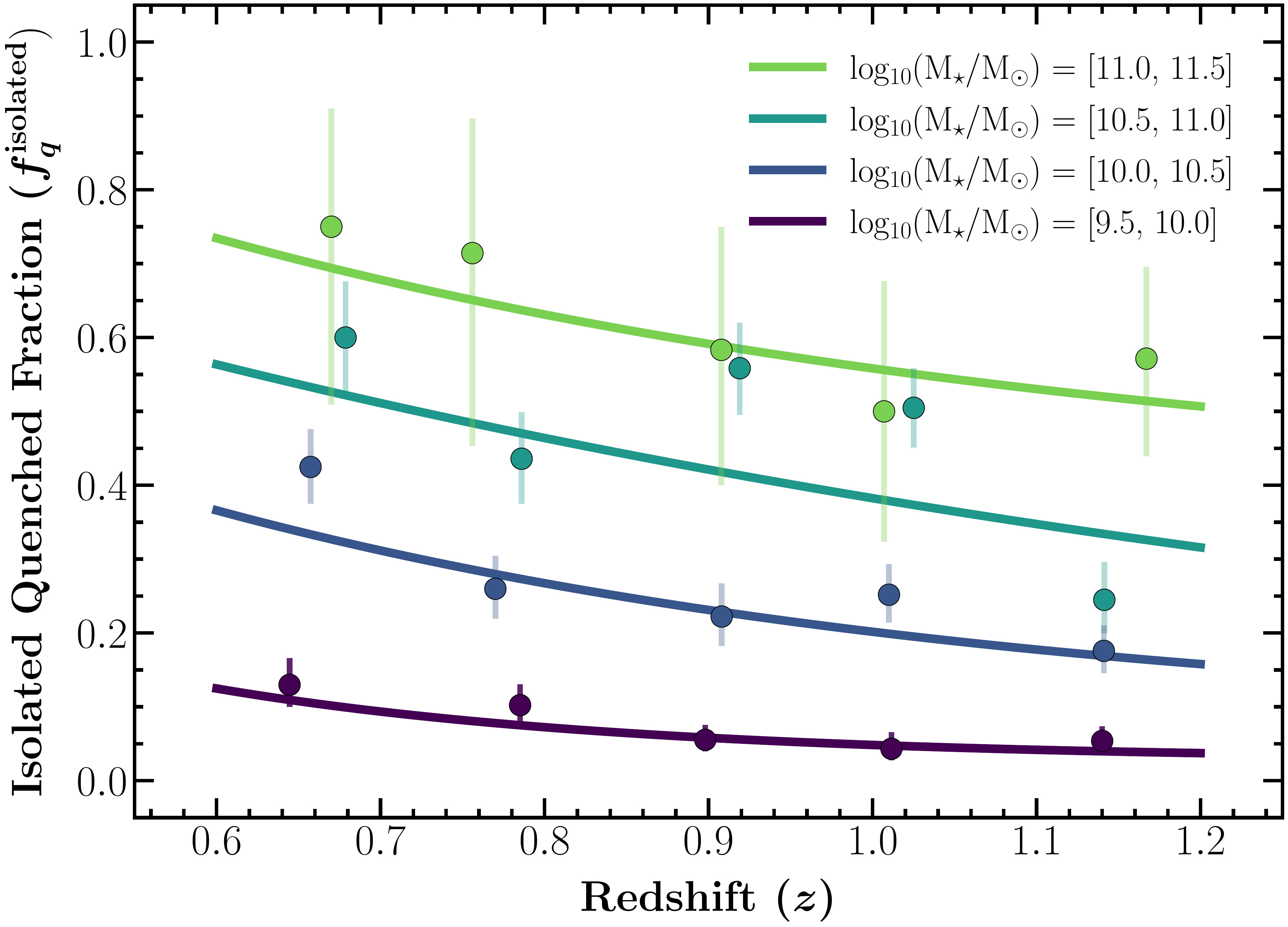}
\caption{Isolated quenched fraction as a function of redshift in bins of stellar mass ranging from $10^{9.5}\msun$ to $10^{11.5}\msun$, derived from galaxies considered relatively isolated (see text for description) in the NBMS catalog. Colored circles represent the observed field quenched fractions in their respective stellar-mass bins, while the curves illustrate fits to the observed results using an exponentially decaying function. Vertical error bars denote the 1-$\sigma$ binomial uncertainties in the quenched fraction.}
\label{fig:fig5}
\end{figure}

\subsection{Contributions from self quenching}
\label{subsec:sim_class}

While environment plays a significant role in the build-up of the passive galaxy population, some galaxies quench through internal mechanisms, such as feedback from star formation and/or accreting supermassive-black holes \citep[e.g.,][]{Oppenheimer06, Croton06}. 
This process, known as \textquote{self-quenching}, acts as a distinct and important contributor to galaxy quenching independent of environment \citep{Peng10}. 
It is therefore reasonable to assume that a fraction of the observed group population may have self-quenched prior to infall.
To account for group members that potentially self-quenched prior to infall, we construct a control sample of isolated galaxies from the NMBS catalog in the EGS (see \S\ref{subsubsec:control-sample}).  
This control sample allows us to measure the quenched fraction of isolated galaxies as a function of mass and redshift, providing a baseline for comparison with the observed group members.
We define isolated galaxies as those located sufficiently far away from massive ($\gt 10^{10.75}~\msun$) galaxies in terms of both projected separation ($\gt 2$ Mpc) and line-of-sight velocity ($\gtrsim$ 5000 $\kms$). 
We determined the threshold for defining massive galaxies by analyzing the stellar mass distribution of the two most massive group members for each group, with $\mstar \sim 10^{10.75}\msun$ being the minimum of the distribution.
Following the UVJ color-color classification scheme highlighted in \S\ref{subsec:classification}, we separate the $\sim 3600$ isolated galaxies into star-forming and quiescent populations. 

Figure~\ref{fig:fig5} illustrates the observed isolated quenched fraction ($f^{\rm isolated}_{{\rm q}}(z, \mstar)$) in bins of redshift and stellar mass. 
We observe a prominent trend between the isolated quenched fraction and stellar mass, indicating that low mass isolated galaxies are predominantly star-forming relative to massive isolated galaxies. 
We find that the observed fractions are robust to variations in both the threshold used to define massive galaxies and the line-of-sight velocity threshold for isolation.
The robustness with respect to the line-of-sight velocity threshold is particularly crucial because, although the isolation condition relies heavily on photometric redshifts, more stringent isolation criteria (e.g., $|\Delta v| > 30,000~\text{km/s}$) do not significantly alter the general trends.

\begin{figure}
\centering
\hspace*{-0.25in}
\includegraphics[width=3.5in]{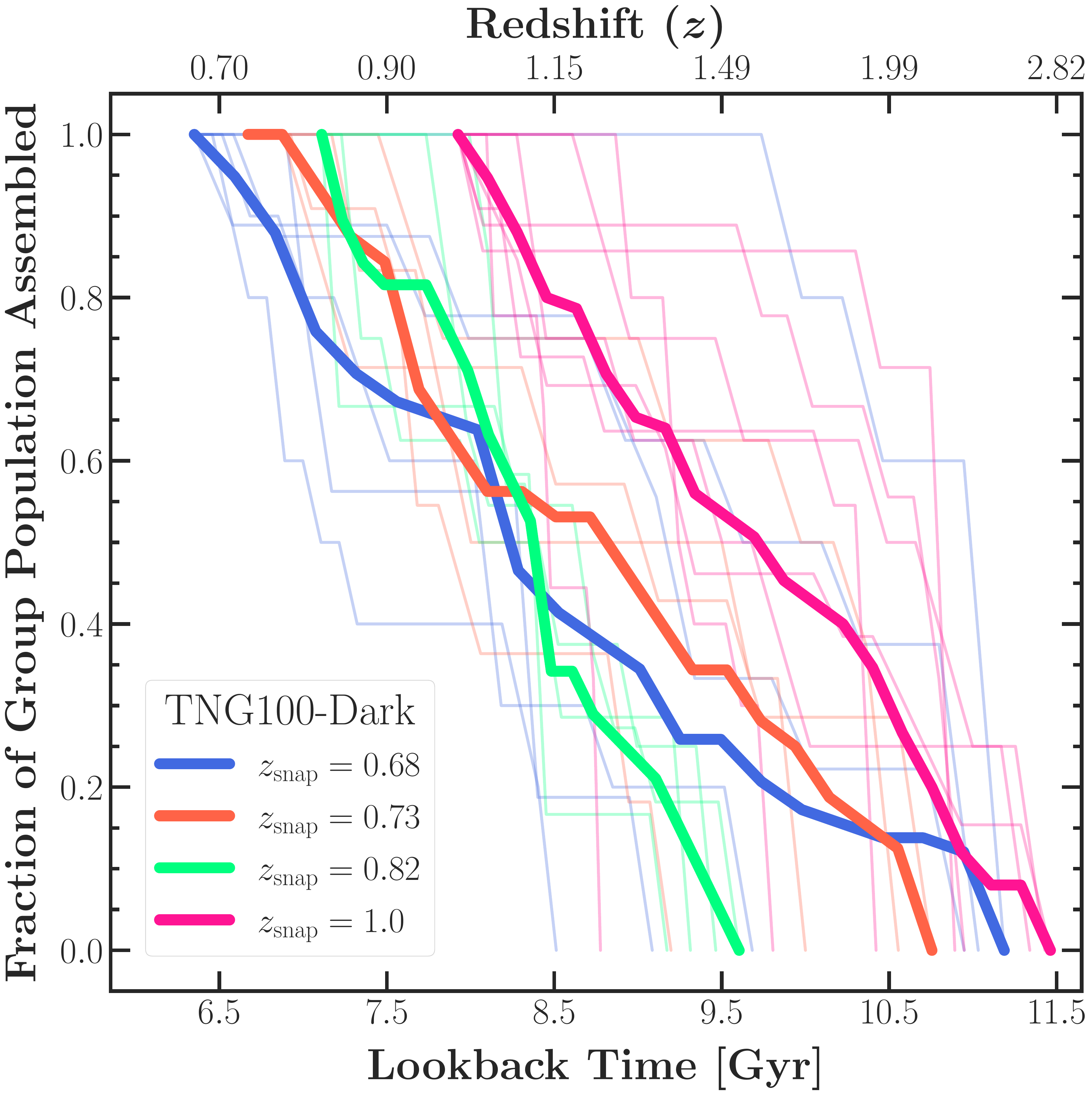}
\caption{Cumulative infall time distribution for the simulated group population. The faded colored lines depict the individual assembly history of group satellites as a function of lookback time (top axis) and redshift (bottom axis). The solid lines represent the combined distribution for all groups at a given snapshot. While individual groups exhibit diversity in their assembly histories, the majority assemble the bulk of their satellite population at redshifts below $z \lt 2$.}
\label{fig:fig6}
\end{figure}

We employ curve fitting techniques to model the isolated quenched fractions, which are then utilized in tandem with the simulated group sample to probabilistically classify the members of simulated groups at the observation redshift. 
That is to say, we use the observed isolated quenched fraction as a function of redshift and stellar mass from Fig.~\ref{fig:fig5} to establish a baseline for the quenched fraction of simulated galaxies that would likely be quenched regardless of their environment.
To achieve this we generate random numbers from a uniform distribution ranging from zero to one and compare them with the expected isolated quenched fraction for galaxies of a given stellar mass and redshift. 
If the generated number exceeds (or falls below) the observed isolated quenched fraction, we classify the simulated galaxy as star-forming (or quenched). 
For instance, as shown in Fig.~\ref{fig:fig5}, we see that an isolated galaxy at $z=1$ with a stellar mass between $10^{10-10.5}~\msun$ has an expected quenched fraction of approximately $20\%$. 
Thus, in this modeling framework, such a galaxy will be classified as quenched only if the randomly generated number is less than $0.2$ (or $20\%$); otherwise, it will be assumed to be star-forming.

We repeat the classification procedure $N$ times (here, $N=10$) to account for the inherent randomness of the procedure. 
Increasing the number of iterations has negligible impact on the results. 
We define the contribution to the group quenched fraction from galaxies that quench independent of environment as the median quenched fraction value across all $N$ iterations.
This process establishes the baseline group quenched fraction for the simulated group population, which is subsequently augmented in bins of stellar mass, group-centric radius, and redshift using our infall-based quenching model to align with observed quenched fractions.


\subsection{Measuring infall times}
\label{subsec:infall_time_distribution}

The second component of our quenching model involves analyzing the infall time distribution of the simulated group population. 
As detailed in \S\ref{subsec:simulated_group_membership} we utilize the TNG100-Dark sublink merger trees to track the positions of group members over time. 
This allows us to pinpoint the infall time ($t_{\rm{infall}}$) for individual simulated satellites, defined as the moment when a subhalo first crosses $R_{200}$ of the group. 
However, due to a median timestep resolution of approximately $100$ Myr between each snapshot, we employ spline interpolation to enhance the temporal resolution to approximately $10$ Myr.

Figure~\ref{fig:fig6} provides a visualization of the infall time distribution, showcasing how the simulation group population assembles over time.
For context, the assembly process is defined from the moment a central galaxy acquires a satellite that remains bound to it until the observed redshift.
This figure highlights the diverse assembly histories of individual groups, while also revealing that, on average, a significant portion (more than $60\%$) of group assembly occurs after $z\lt2$.

\begin{figure*}
\centering
\hspace*{-0.1in}
\includegraphics[width=6.5in]{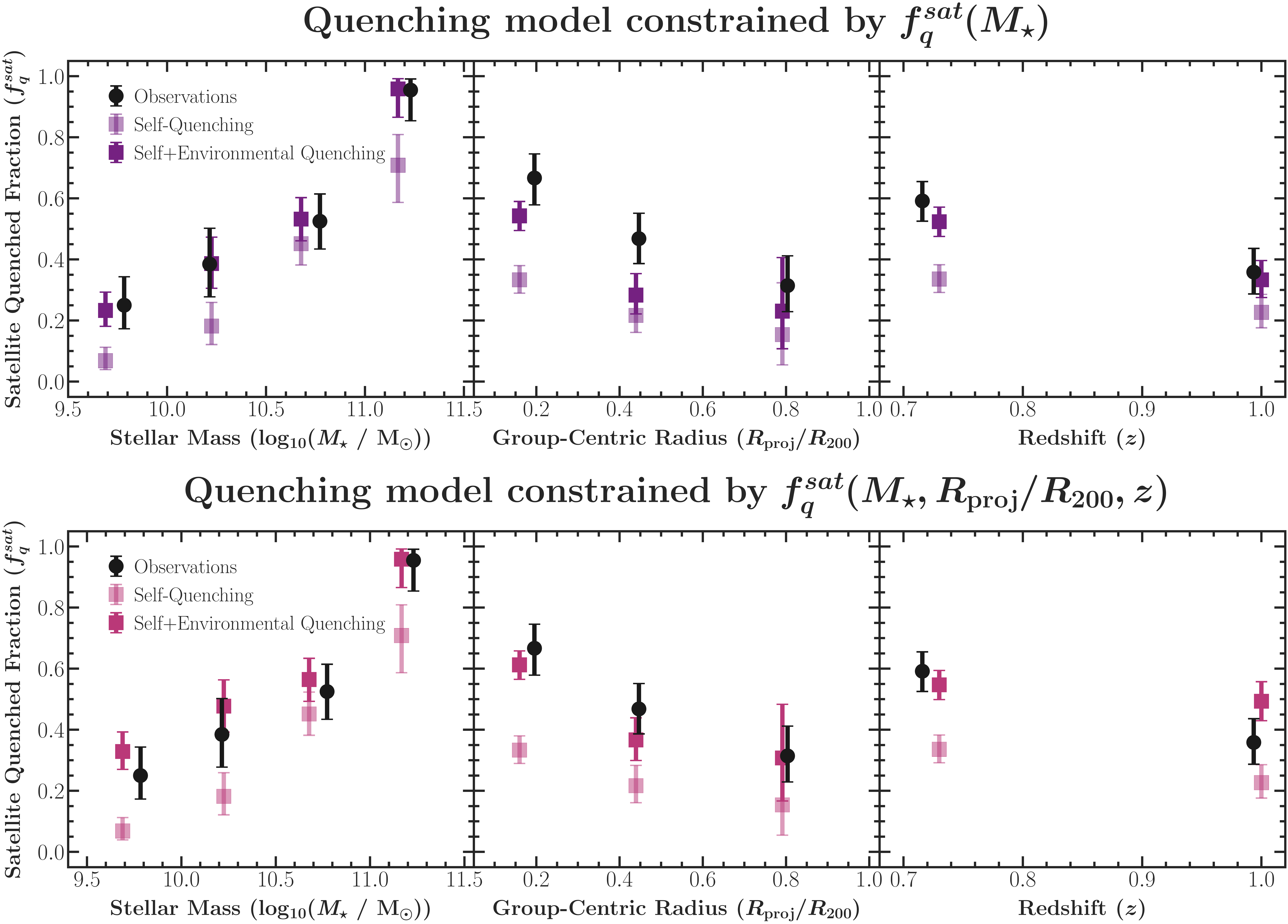}
\caption{Satellite quenched fraction as a function of satellite stellar mass (left panel), projected group-centric radius (middle panel), and redshift (right panel). Black circles represent observed quenched fractions, while light purple and magenta squares denote modeled quenched fractions for satellites that quenched independently of the group environment. Dark purple and magenta squares represent the best-fit quenched fractions from the quenching model for the total satellite population, including galaxies that quenched both independently and as a result of the group environment. The top and bottom panels show results where the quenching model is constrained by $f_{q}^{\mathrm{sat}}(\mstar)$ and $f_{q}^{\mathrm{sat}}(\mstar, R_{\mathrm{proj}}/\rtwo, z)$, respectively.}
\label{fig:fig7}
\end{figure*}

\begin{figure*}
\centering
\includegraphics[width=5.5in]{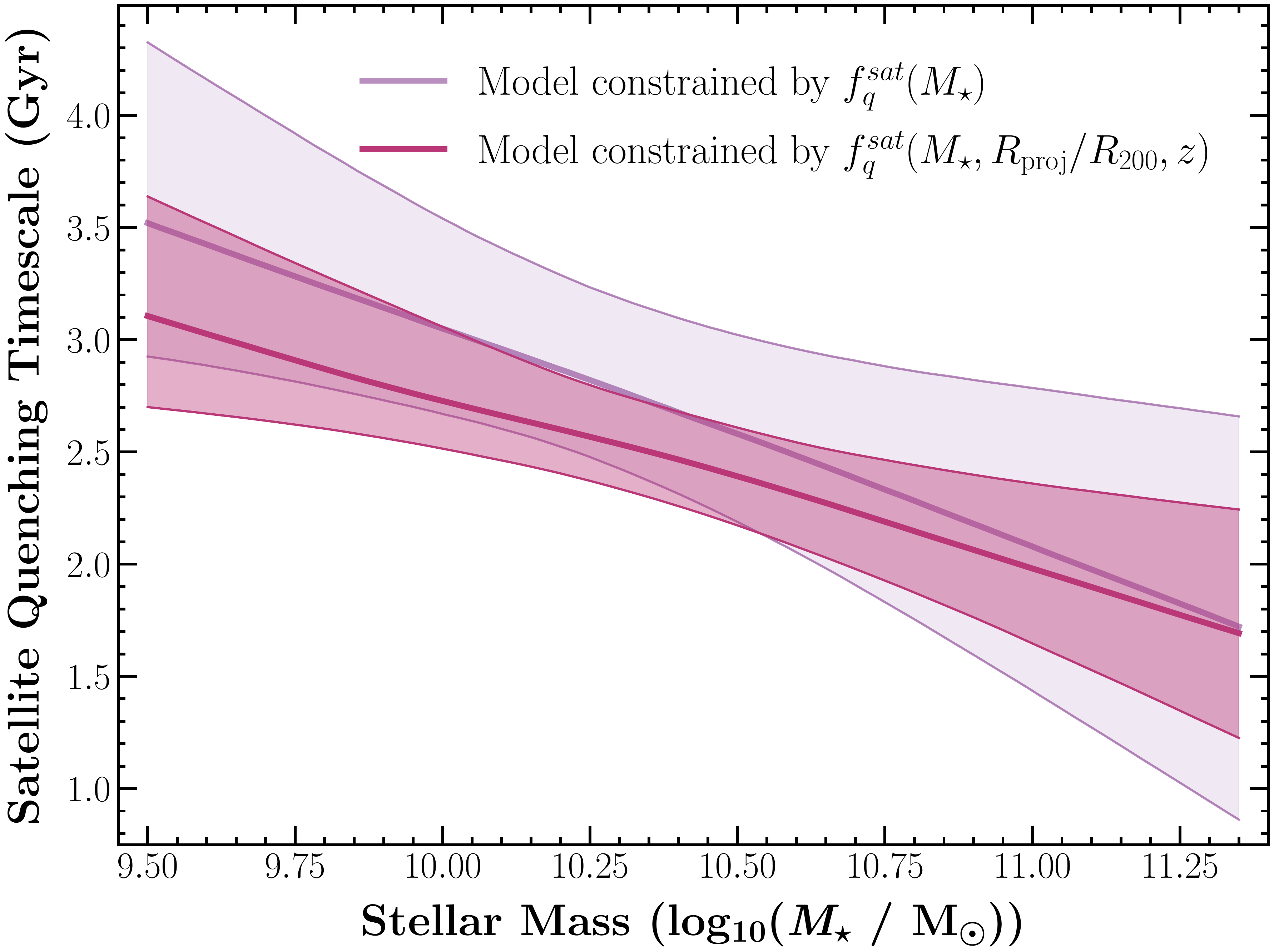}
\caption{Satellite quenching timescale as a function of stellar mass, with median values and 1-sigma uncertainties for two scenarios. The purple line shows the timescale constrained by the observed trend of quenched fraction with stellar mass ($f_{q}^{\mathrm{sat}}(\mstar)$), while the magenta line includes constraints from stellar mass, group-centric radius, and redshift ($f_{q}^{\mathrm{sat}}(\mstar, R_{\mathrm{proj}}/\rtwo, z)$). In both cases, a mass-dependent quenching timescale is evident, decreasing from $\sim3-3.5$ Gyr at $10^{9.5}\msun$ to $\sim2.5$ Gyr at $10^{10.5}\msun$.}
\label{fig:fig8}
\end{figure*}

\subsection{Constraining quenching timescales}
\label{subsec:quenching_timescales}

While environmental quenching is a complex process that depends on more than a single parameter, constraining the timescale upon which environmental quenching proceeds, denoted $\tau_{\mathrm{quench}}$, has proven valuable in isolating dominant quenching mechanisms in groups and clusters \citep[e.g.,][]{DeLucia12, CW14, wheeler14, Fham15, Fossati17, Foltz18}. In general, studies that constrain $\tau_{\mathrm{quench}}$ typically define the onset of environmental quenching relative to the moment a less massive subhalo becomes a member of a more massive subhalo. This can be defined relative to the instance when a subhalo crosses the virial radius of any subhalo more massive than itself \emph{or} the instance when a subhalo first crosses the virial radius of its final host. The former typically helps constrain the role of group pre-processing in building up the quiescent population of galaxy clusters. The latter, which is the method utilized in this work, is used to constrain the timescale upon which dense environments suppress star formation. 

In this study, we use the assembly histories from TNG100-Dark to constrain $\tau_{\mathrm{quench}}$. First, we establish $t_{\rm{max}}$, representing the maximum time available for a subhalo to quench, defined as the duration between its first infall and the observed snapshot ($t_{\rm{max}}= t_{\rm{infall}} - t_{\rm{obs}}$). Additionally, we parameterize $\tau_{\mathrm{quench}}$ as a linear function of stellar mass with parameters $m$ (slope) and $b$ (y-intercept):

\begin{equation}
\label{eq:tau_quench}
\tau_{\rm{quench}} = m*\log_{10}(\mstar/\msun) + b.
\end{equation}.

Therefore, the condition for a simulated group member to quench is given by $\tau_{\mathrm{quench}} \leq t_{\mathrm{max}}$. Thus, after accounting for the sub-population of simulated galaxies that quench independently of the group environment (see \S\ref{subsec:sim_class}), we can vary $\tau_{\mathrm{quench}}$ to generate a range of modeled quenched fractions. To efficiently explore this parameter space, we employ Markov Chain Monte Carlo (MCMC) sampling techniques. This enables us to identify $\tau_{\mathrm{quench}}$ values that best align with both modeled and observed quenched fractions.

\section{Results}
\label{sec:Results}

\subsection{Satellite quenched fractions}
\label{subsec:fq_results}

We stack our observed group populations, a total of 124 satellites with stellar masses greater than $\mstar \gt 10^{9.5}~\msun$ across 11 X-ray selected groups. 
Using this combined sample, we measure the satellite quenched fraction as a function of satellite stellar mass, group-centric radius (measured from the luminosity-weighted centroid), and redshift. 
We find that the fraction of quenched galaxies increases from $\sim20$ percent in the lowest mass bin to around $95$ percent in the highest mass bin (see Fig.~\ref{fig:fig7}). 
This trend is roughly consistent with results from previous group quenching studies that combine spectroscopically and photometrically confirmed members to measure the satellite quiescent fraction in groups at $z\sim1$ down to 10$^{9.5}~\msun$ \citep[e.g.,][]{Balogh16, Fossati17}. 
Compared to measurements from group studies at $z\sim0$, the satellite quenched fractions measured here are roughly two times lower at low masses ($\mstar \sim 10^{9.5}~\msun$) and nearly unchanged at high masses ($\mstar \gt 10^{11.0}~\msun$) \citep[e.g.,][]{Baxter21, Meng23}.
Regarding the dependence on redshift, we observe an increase in the quenched fraction from $35$ percent at $z\sim1$ to $60$ percent at $z\sim0.7$. 
Lastly, we find that the fraction of quiescent galaxies increases with decreasing group-centric radius. 
This is somewhat to be expected given that the quenched fraction is measured relative to the luminosity-weighted centroid, but this trend still holds (albeit not as steep) if measured relative to the position of peak X-ray emission. 

Environmental quenching studies often use the observed relationship between the quenched fraction and a single parameter, such as satellite stellar mass or host-centric radius, to constrain the satellite quenching timescale \citep[e.g.,][]{Wetzel13, wheeler14, Balogh16, Baxter22}.
This approach is practical because it captures the intended quenched fraction trend as a function of the chosen parameter.
 However, as illustrated in the top row of Fig.~\ref{fig:fig7}, this method can struggle to capture the observed variations in the quenched fraction with other parameters, like group-centric radius.
A more comprehensive approach is to constrain the model based on trends in stellar mass, group-centric radius, and redshift.
While this method prioritizes capturing the interplay between all three parameters, sacrificing a perfect match with the quenched fraction versus stellar mass trend (as shown in the bottom row of Fig.~\ref{fig:fig7}), it yields improved overall alignment with all three observed trends and provides tighter constraints on the inferred quenching timescale, as elaborated below.

\subsection{Satellite quenching timescales}
\label{subsec:timescale_results}

Fig.~\ref{fig:fig8} depicts the satellite quenching timescales as a function of satellite stellar mass inferred from this investigation.
We present results for two scenarios: the quenching timescale constrained solely by the observed quenched fraction trend with stellar mass ($f_{q}^{\mathrm{sat}}(\mstar)$) depicted by the purple line, and constrained by quenched fraction trends with stellar mass, group-centric radius, and redshift ($f_{q}^{\mathrm{sat}}(\mstar, R_{\mathrm{proj}}/\rtwo, z)$) shown by the magenta line in Fig.~\ref{fig:fig8}.
We find that by jointly constraining the quenching model against multiple quenched fraction trends, we achieve tighter constraints on the quenching timescale across the entire range of investigated stellar masses.
Nevertheless, in both cases, we infer a mildly mass-dependent quenching timescale in which higher-mass satellites quench more rapidly relative to lower-mass counterparts. 
Specifically, the quenching timescales that we infer decrease with increasing satellite stellar mass, ranging from approximately $3-3.5~\mathrm{Gyr}$ at $10^{9.5}~\msun$ to $\sim2.5~\mathrm{Gyr}$ at $10^{10.5}~\msun$.
In the following section we discuss the ramifications of these timescales and how they compare with estimates from previous studies in the literature.

\section{Discussion}
\label{sec:Discussion}

\subsection{The dominant quenching pathway in galaxy groups}
\label{subsec:quenching_mech}

\begin{figure*}
\centering
\includegraphics[width=6.5in]{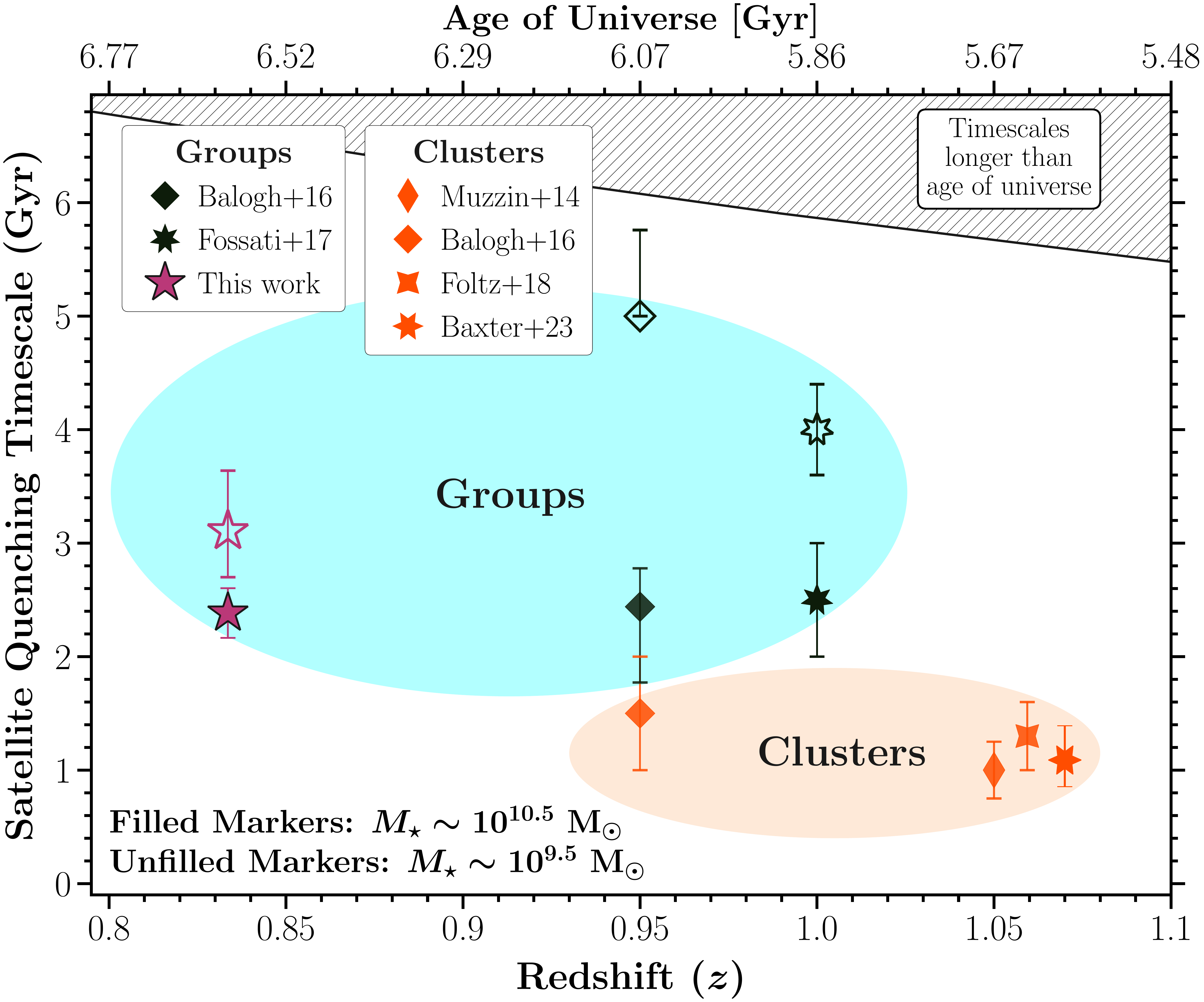}
\caption{Comparison of satellite quenching timescales from the literature at $z\sim1$, shown for groups (black markers) and clusters (orange markers). The results from this work are denoted by magenta stars. Filled and unfilled markers depict high-mass ($\gtrsim10^{10.5}\msun$) and low-mass ($\gtrsim10^{9.5}\msun$) satellites, respectively. Notably, groups exhibit longer quenching timescales for similar-mass galaxies compared to clusters. Additionally, our study finds shorter quenching timescales for lower-mass group galaxies relative to photometry-based estimates, suggesting that low-mass satellites quench faster than previously thought.}
\label{fig:fig9}
\end{figure*}

Fundamentally, the quenching of star formation requires two ingredients. First, the inflow of cold gas (i.e. fuel for star formation) onto the galaxy must be halted. Second, the remaining star-forming gas must be consumed, ejected, or stripped away from the disk. Regarding the former, simulations show that galaxies within host halos with masses above $\mhalo \gtrsim 10^{13}$ are able to halt the inflow of cold gas streams from the intergalactic medium, with more massive host halos more efficiently cutting off accretion \citep[e.g.,][]{vdV17}. Thus, in group  environments this is thought to be an effective method to halt the accretion of pristine cold gas, though this process is less efficient in groups than in clusters. Regarding the latter, the consumption of cold gas can be framed in terms of the depletion timescale ($\tau_{\mathrm{dep}} \equiv M_{\mathrm{gas}}/SFR$) of molecular gas or atomic and molecular gas. This is assumed to also include the stripping of the extended hot gas halo surrounding a galaxy -- which serves as an additional source of star-forming gas after a satellite is separated from gas inflows from the IGM -- via a process dubbed \textquote{strangulation} \citep{BM00}. Additionally, the interactions between a galaxy and its host halo provides several opportunities to strip star-forming gas either via dynamical interactions (e.g., RPS) or gravitational interactions (e.g, tidal stripping). The timescales for these mechanisms typically scale like the crossing -- or dynamical -- time ($\tau_{\mathrm{dyn}} \equiv R/V$).

The quenching timescales inferred in this analysis are longer than both the typical crossing time ($\lesssim 1$ Gyr) and the total cold gas (H{\scriptsize I}+H$_{2}$) depletion timescale ($\lesssim2$ Gyr at  $z\sim1$ for galaxies with $\mstar\sim10^{10}~\msun$, based on estimates from \citet{Popping15}). While direct stripping of star-forming gas through ram-pressure stripping is believed to occur in group environments \citep[e.g.,][]{Sengupta07}, it is expected to be relatively inefficient due to lower densities and velocities compared to cluster environments \citep[for a recent review, see][]{Cortese21}. This, coupled with the relatively long quenching timescales, suggests that direct stripping events are unlikely to be the dominant drivers of star formation suppression in groups.

A possible scenario highlighted in \citet{McGee14, Fossati17} to explain timescales that exceed the expected total cold gas depletion timescale is that group environments are relatively inefficient at removing the hot gas halo surrounding infalling galaxies. In this scenario the multi-phase gas in the circumgalactic medium (CGM) of the satellite can potentially cool onto the galaxy, replenishing the supply of star-forming gas and sustaining continued star formation. Thus the relatively long timescales inferred from this analysis are consistent with the idea that quenching in galaxy groups is mainly driven by gas exhaustion, with the caveat that quenching timescales may exceed estimated depletion timescales due to CGM gas recycling.

\begin{figure*}
\centering
\includegraphics[width=6.8in]{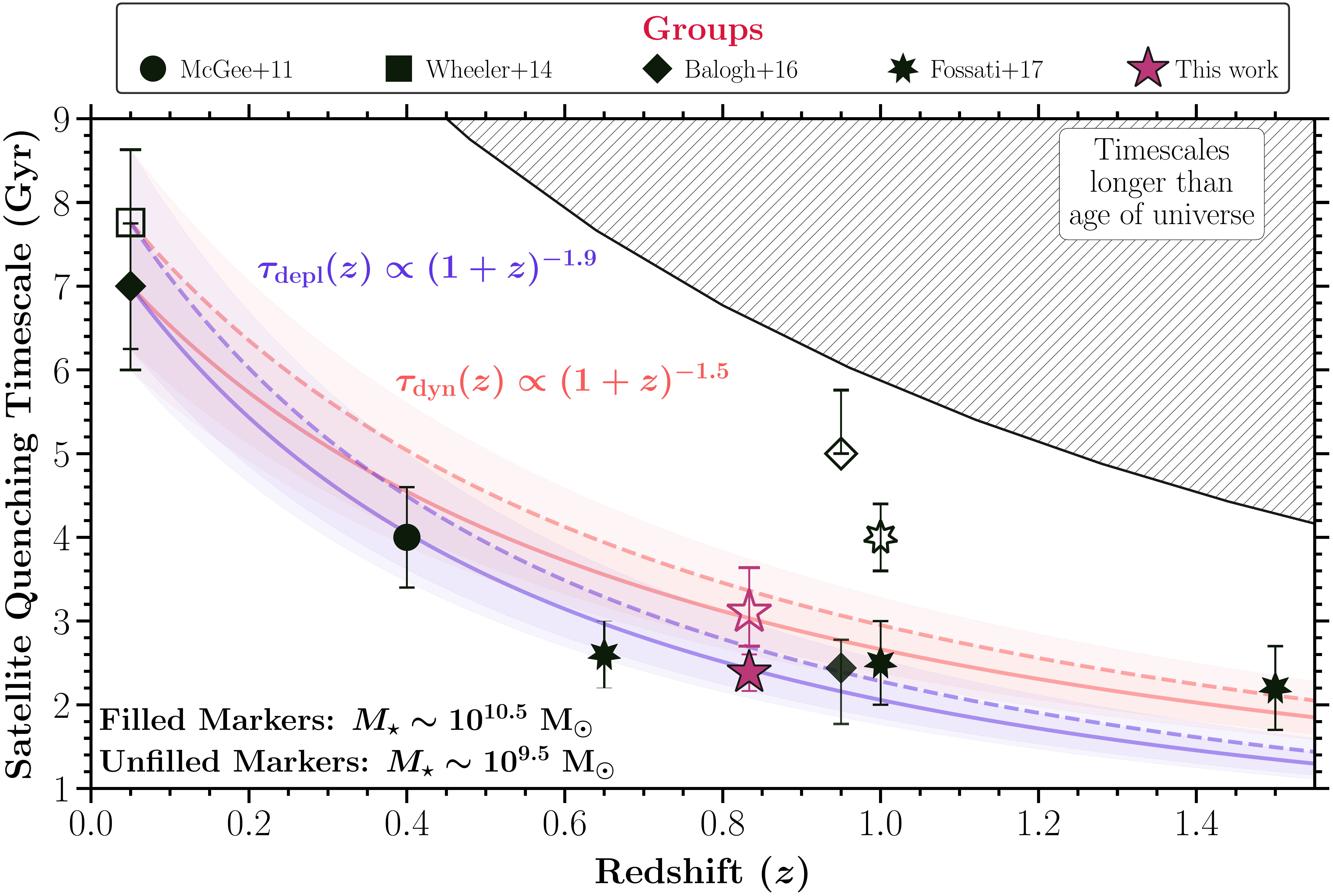}
\caption{Satellite quenching timescale versus redshift for galaxy groups. Filled and unfilled markers represent high-mass ($\gtrsim10^{10.5}\msun$) and low-mass ($\gtrsim10^{9.5}\msun$) satellites, with solid and dashed lines also corresponding to these respective masses. The orange lines illustrate the evolution of dynamical time ($\propto (1+z)^{-1.5}$), whereas the purple lines depict the evolution of the total cold gas (H{\scriptsize I}+H$_{2}$) depletion time ($\propto (1+z)^{-1.9}$), derived from \citet{Popping15}. Both trends are normalized by the quenching timescales at $z\sim0$ from \citet{Balogh16} ($7.0\substack{+0.75 \\ -0.75}$ Gyr at $\gtrsim10^{10.5}\msun$) and \citet{wheeler14} ($7.8\substack{+0.87 \\ -1.76}$ Gyr at $\gtrsim10^{9.5}\msun$). The timescale constraints at low and high masses from this work are consistent with both evolutionary scenarios, suggesting that quenching in galaxy groups over the past 7 billion years may be driven either by the stripping of cold gas or by its passive consumption in the absence of cosmic web accretion.}
\label{fig:fig10}
\end{figure*}

\subsection{The efficiency of quenching in groups vs. clusters}
\label{subsec:grp_vs_cluster_discussion}

In Figure~\ref{fig:fig9} we compare satellite quenching timescales reported in the literature, focusing on studies of galaxy groups ($\mtwo = 10^{13-14}~\msun$) and galaxy clusters ($\mtwo = 10^{14-15}~\msun$) at $z\sim1$. Our results (purple stars) reaffirm previous findings indicating that galaxies with $\mstar \sim 10^{10.5}~\msun$ in group environments \citep[e.g.,][]{Balogh16, Fossati17} quench on timescales of $\sim2.5$ Gyr. These timescales are longer than those inferred from cluster studies, which generally find that galaxies with $\mstar \sim 10^{10.5}~\msun$ quench on timescales of $1-1.5$ Gyr \citep[e.g.,][]{Muzzin14, Balogh16, Foltz18, Baxter23}. As shown in \citet{Baxter22}  the quenching timescales for clusters at $z\sim1$ is consistent with estimates of the total cold gas depletion timescale at this epoch, suggesting that starvation plays a dominant role in suppressing star formation within cluster environments. Therefore, if the exhaustion of cold gas in the absence of cosmological accretion is also the dominant driver of quenching in groups, then the relatively longer quenching timescales in group environments can be attributed to their retention of the circumgalactic medium (CGM), which serves as a source of additional star-forming gas. In contrast, clusters, due to their high densities and temperatures, are more likely to strip the CGM from infalling satellites, resulting in quenching timescales comparable to the total gas depletion timescale.

\subsection{Redshift evolution of satellite quenching in group environments}
\label{subsec:z_evo_discussion}

The environmental quenching timescale inferred for low-mass ($\mstar \sim 10^{9.5}~\msun$) satellites in groups at $z\sim0$ (approximately 8 Gyr) conflicts with the age of the universe as early as $z\sim0.6$.  
This constraint necessitates an increase in the efficiency with which low-mass group satellites quench as redshift increases. 
Studies at $z\sim1$ that utilize photometrically-selected group members at these masses indeed infer quenching timescales shorter than those inferred at $z\sim0$ \citep{Balogh16, Fossati17}.
The physics driving the redshift evolution of the quenching timescale remains unclear, but several investigations find that it is possibly related to gas stripping events, as this evolution is consistent with the evolution of the dynamical time \citep{McGee14, Foltz18}.
To explore this, Figure~\ref{fig:fig10} presents the quenching timescale as a function of redshift for galaxies in groups, with this investigation providing two new constraints at $z\sim0.8$ for low-mass ($\mstar=10^{9.5}\msun$) and high-mass ($\mstar=10^{10.5}\msun$) satellites in groups.
When placed into the larger context with other studies from the literature, we find results consistent with the picture in which the quenching timescales undergo strong evolution with redshift — nearly doubling over the last 7-8 Gyr, implying that groups with similar halo masses at $z\sim0$ and $z\sim1$ quench their satellite populations at vastly different timescales.
As illustrated by the orange bands in Fig.~\ref{fig:fig10}, our investigation finds that the evolution of the quenching timescale roughly follows the evolution in the dynamical time, i.e., $\tau_{\mathrm{quench}}(z) \propto (1+z)^{-1.5}$, when normalized to the quenching timescale of galaxy groups at $z\sim0$. 
This finding aligns with previous studies on the redshift evolution of environmental quenching timescales \citep[e.g.,][]{McGee14, Foltz18}, suggesting that changes in quenching efficiency result from the dynamical evolution of groups with redshift.
However, as indicated by the purple bands in Fig.~\ref{fig:fig10}, we also find that the redshift evolution of the environmental quenching timescale is consistent with the evolution of the total cold gas (H{\scriptsize I} + H$_{2}$) depletion time derived from \citet{Popping15}, which scales as $\propto (1+z)^{-1.9}$.
This would instead imply that the redshift evolution of environmental quenching in group environments is driven by a decrease in the efficiency of quenching via starvation with decreasing redshift, as opposed to gas stripping events.
This scenario is supported by both this investigation and prior studies at $z\sim0$, which consistently identify starvation as the primary driver of environmental quenching in group environments \citep{VanDenBosch08, Wetzel13, wheeler14, Fham16}.
While stripping likely plays a role in quenching the group satellite population, the relatively long quenching timescales in group environments suggest that less efficient processes like starvation and strangulation dominate.

\section{Summary and Conclusions}
\label{sec:Conclusion}

Using Keck/DEIMOS, we conduct follow-up spectroscopy targeting faint ($R \lesssim 25.5$) satellite candidates around 11 X-ray selected groups in the EGS, covering a redshift range of $0.66 < z < 1.02$.
By stacking confirmed group members, we measure the satellite quenched fraction as a function of stellar mass, group-centric radius, and redshift, to a stellar mass limit of approximately $10^{9.5}~\msun$.

These observed trends are used to constrain an infall-based quenching model, parameterized by the satellite quenching timescale ($\tau_{\rm quench}$), representing the duration a satellite remains star forming after becoming a group member.
This timescale, which is allowed to vary with stellar mass, is estimated using the assembly histories of 22 mock groups from the TNG100-Dark simulation, selected to mirror the redshift and halo mass distribution of the 11 observed groups.
Importantly, before estimating $\tau_{\rm quench}$, the model calibrates against the observed coeval quenched fraction of isolated galaxies, establishing a baseline quenched fraction for group members that quenched independently of their environment.
Once this baseline is established, additional galaxies are allowed to quench based on their group membership duration.
This modeling procedure, conducted in a Bayesian framework, constrains the quenching timescale that best matches the modeled quenched fractions with the observations.
The primary findings from this analysis are as follows:

\begin{enumerate}

\item At lower masses ($\mstar = 10^{9.5}~\msun$) we infer quenching timescales of $3.1\substack{+0.5 \\ -0.4}~\mathrm{Gyr}$, while at higher masses ($\mstar = 10^{10.5}~\msun$) we find  shorter quenching timescales of $2.4\substack{+0.2 \\ -0.2}~\mathrm{Gyr}$, consistent with previous estimates from the literature.
\item These relatively long timescales are consistent with starvation as the dominant quenching mechanism, provided the multi-phase circumgalactic medium remains intact post-infall onto the group. This condition allows for hot diffuse gas to gradually cool onto the galaxy from the CGM, thereby enabling prolonged star formation. Given that the observed timescales surpass the estimated total cold gas depletion timescale at this epoch, the preservation of the circumgalactic medium emerges as a critical factor in this study.

\item We observe a weaker dependence on the quenching timescale with satellite stellar mass compared to previous studies. Our findings for $\tau_{\rm quench}$ at $\mstar = 10^{10.5}\msun$ agree with the existing literature. Therefore, the shallower trend is predominantly driven by quenching timescales at $\mstar = 10^{9.5}\msun$, which are shorter than prior photometry-based estimates from the literature. This suggests that quenching in the low-mass regime is more efficient than previously thought.

\item When comparing quenching timescales at $z\sim1$ from the literature for groups and clusters, we find that high mass ($\gtrsim10^{10.5}~\msun$) cluster satellites quench on timescales of $1-1.5$ Gyr, while high mass ($\gtrsim10^{10.5}~\msun$) group satellites have slightly longer timescales of $\sim2.5$ Gyr. This mild halo mass dependence is attributed to clusters more efficiently removing the CGM of satellites, thereby eliminating a potential source of additional star-forming gas post-infall.
\item We place novel constraints on the quenching timescale at $z\sim0.8$ for group satellites with $\mstar = 10^{9.5}\msun$ and $\mstar = 10^{10.5}\msun$. By exploring the redshift dependence of the satellite quenching timescale in groups, we find that quenching efficiency increases with increasing redshift. Our results indicate that the quenching timescale evolves roughly like the dynamical time ($\propto(1+z)^{-1.5}$) and is consistent with previous studies \citep[e.g.,][]{Balogh16, Foltz18} and the total cold gas depletion time from \citet{Popping15} ($\propto(1+z)^{-1.9}$). This suggests that the near doubling of the environmental quenching timescale over the last 7-8 Gyr could be due to the dynamical evolution of groups or a decrease in quenching efficiency via starvation with decreasing redshift. While stripping likely plays a role, the relatively long quenching timescale in groups at $z\sim0$ and $z\sim1$ suggests that starvation is the primary quenching pathway.
\end{enumerate}


The results from this study represent only the tip of the iceberg in understanding low-mass satellite quenching in overdense regions at early times. Since these data were collected, deeper photometric surveys in legacy fields, such as COSMOS2020 \citep{Weaver23}, have become publicly available. Additionally, data from JWST are enabling the spectroscopic confirmation of low-mass galaxies down to $\mstar \sim 10^{8}~\msun$ at cosmic noon ($2 < z < 3$) \citep{Cutler24}. While not specifically focused on environment, these studies identify distinct populations of quiescent galaxies at low and high stellar masses, suggesting different quenching pathways, similar to what is observed at later times. Looking ahead, these new facilities and datasets will enable the selection of even fainter satellite candidates around spectroscopically-confirmed groups and clusters at early times. This will undoubtedly deepen our understanding of environmental quenching in low-mass galaxies and its evolution across cosmic time.

\section*{Acknowledgements} 

DCB is supported by an NSF Astronomy and Astrophysics Postdoctoral Fellowship under award AST-2303800. DCB is also supported by the UC Chancellor's Postdoctoral Fellowship.
This study makes use of data from the NEWFIRM Medium-Band Survey, a multi-wavelength survey conducted with the NEWFIRM instrument at the KPNO, supported in part by the NSF and NASA.
Most of the data presented herein were obtained at Keck Observatory, which is a private 501(c)3 non-profit organization operated as a scientific partnership among the California Institute of Technology, the University of California, and the National Aeronautics and Space Administration. The Observatory was made possible by the generous financial support of the W. M. Keck Foundation.
The authors wish to recognize and acknowledge the very significant cultural role and reverence that the summit of Maunakea has always had within the Native Hawaiian community. We are most fortunate to have the opportunity to conduct observations from this mountain.
The authors also thank the anonymous referee for their critical assessment of our work and for providing valuable comments that have enhanced the clarity of the paper.

\vspace{5mm}
\facilities{Keck:II (DEIMOS)}

\vspace{5mm}
\software{\texttt{astropy} \citep{Astropy13, Astropy18, Astropy22}, \texttt{BAGPIPES} \citep{Carnall18},  \texttt{emcee} \citep{Foreman13}, \texttt{iPython} \citep{iPython}, \texttt{pandas} \citep{reback2020pandas}, \texttt{matplotlib} \citep{matplotlib}, \texttt{NumPy} \citep{numpy}, \texttt{spec2d} \citep{Cooper12} }

\bibliography{citations}{}
\bibliographystyle{aasjournal}

\end{document}